\documentclass[fleqn,usenatbib]{mnras}

\usepackage{newtxtext,newtxmath}
\usepackage[T1]{fontenc}
\usepackage{ae,aecompl}
\usepackage{graphicx}
\graphicspath{{Figures/}}
\usepackage{tabularx}
\usepackage{amsmath}
\usepackage{amssymb}
\usepackage{gensymb}
\usepackage{wasysym}
\usepackage{color}

\newcolumntype{Y}[1]{>{\centering\arraybackslash}p{#1}}
\newcolumntype{Z}[1]{>{\raggedright\arraybackslash}p{#1}}

\title[\textsc{Mock-X} mass bias]{Characterizing hydrostatic mass bias with \textsc{Mock-X}}

\author[D. J. Barnes et al.]{David J. Barnes$^{1}$\thanks{E-mail: djbarnes@mit.edu}, Mark Vogelsberger$^{1}$, Francesca A. Pearce$^{2}$, Ana-Roxana Pop$^{3}$,\newauthor Rahul Kannan$^{3}$, Kaili Cao$^{1}$, Scott T. Kay$^{2}$, Lars Hernquist$^{3}$
\\
$^1${Department of Physics, Kavli Institute for Astrophysics and Space Research, Massachusetts Institute of Technology, Cambridge, MA 02139, USA}\\
$^2$Jodrell Bank Centre for Astrophysics, Department of Physics and Astronomy, The University of Manchester, Manchester M13 9PL, UK\\
$^3${Center for Astrophysics | Harvard \& Smithsonian, 60 Garden Street, Cambridge, MA 02138}\\
}

\date{Accepted XXX. Received YYY; in original form ZZZ}

\pubyear{2018}

\begin{document}
\label{firstpage}
\pagerange{\pageref{firstpage}--\pageref{lastpage}}
\maketitle

\begin{abstract}
Surveys in the next decade will deliver large samples of galaxy clusters that transform our understanding of their formation.
Cluster astrophysics and cosmology studies will become systematics limited with samples of this magnitude.
With known properties, hydrodynamical simulations of clusters provide a vital resource for investigating potential systematics.
However, this is only realized if we compare simulations to observations in the correct way.
Here we introduce the \textsc{Mock-X} analysis framework, a multiwavelength tool that generates synthetic images from cosmological simulations and derives halo properties via observational methods.
We detail our methods for generating optical, Compton-$y$ and X-ray images.
Outlining our synthetic X-ray image analysis method, we demonstrate the capabilities of the framework by exploring hydrostatic mass bias for the IllustrisTNG, BAHAMAS and MACSIS simulations. 
Using simulation derived profiles we find an approximately constant bias $b\approx0.13$ with cluster mass, independent of hydrodynamical method or subgrid physics.
However, the hydrostatic bias derived from synthetic observations is mass-dependent, increasing to $b=0.3$ for the most massive clusters.
This result is driven by a single temperature fit to a spectrum produced by gas with a wide temperature distribution in quasi-pressure equilibrium.
The spectroscopic temperature and mass estimate are biased low by cooler gas dominating the emission, due to its quadratic density dependence.
The bias and the scatter in estimated mass remain independent of the numerical method and subgrid physics.
Our results are consistent with current observations and future surveys will contain sufficient samples of massive clusters to confirm the mass dependence of the hydrostatic bias.
\end{abstract}
\begin{keywords}
methods: numerical -- galaxies: clusters: general -- galaxies: clusters: intracluster medium -- X-rays: galaxies: clusters
\end{keywords}

\section{Introduction}
\label{sec:intro}
Galaxy clusters form from the largest amplitude fluctuations present in the early Universe.
Growing hierarchically over cosmic time as gravity draws in gas, stars, dark matter and other collapsed structures, galaxy clusters are the most massive collapsed objects we encounter at the current epoch.
The distribution of galaxy clusters observed as a function of mass and redshift depends strongly on the initial spectrum and growth of the primordial fluctuations \citep[e.g.][]{Davis1985,Peacock1985,Bardeen1986}.
Therefore, clusters have the potential to place stringent constraints on the fundamental cosmological parameters that describe the Universe \citep[e.g.][]{Allen2011,Kravtsov2012}, including the nature of dark energy \citep[e.g.][]{Weinberg2013}.
However, a prerequisite for cluster cosmology studies is well characterized observable-mass relations \citep[e.g.][]{Reiprich2002,Vikhlinin2003,Giodini2013}.
Given that dark matter comprises $\sim85$ per cent of most clusters, mass estimates are difficult and require very high-quality data.
Additionally, as galaxy clusters collapse the galaxies residing within them grow and evolve via a range of astrophysical processes, such as radiative cooling, star formation, supernovae and the energetic outbursts of supermassive black holes.
These processes continually shape the cluster's baryonic components, making many observables a complex interplay of both cosmology and astrophysics.
Therefore, for cluster cosmology studies a detailed understanding of both the mass estimate and observed properties are critical.

Galaxy clusters have proven to be an invaluable tool in placing relatively competitive constraints on fundamental cosmological parameters \citep[e.g.][]{Mantz2014,deHaan2016,Bocquet2019}, which includes dark energy, the summed neutrino masses \citep[e.g.][]{Mantz2015wtg,Madhavacheril2017} and modifications to gravity \citep[e.g.][]{Okabe2013,Wilcox2015}.
Additionally, the constraints provided by clusters are often orthogonal to those produced via other methods.
Due to the high-quality data required for reliable mass estimation, many studies are currently limited by statistical errors due to small sample sizes.
However, the current decade will see a transformation in galaxy cluster observations.
Surveys from facilities such as \textit{Euclid} \citep{Laureijs2011}, LSST \citep{LSST2009}, \textit{e-Rosita} \citep{Merloni2012}, SPT-3G \citep{Benson2014} and the Simons Observatory \citep{SOC2019} will yield samples with $>10^{5}$ objects, orders of magnitude larger than currently available.
Combined with extensive follow-up programs, these surveys will provide a detailed, multiwavelength $10\,\mathrm{Gyr}$ picture of the growth and evolution of clusters.
In this new regime of precision cluster cosmology, systematic uncertainties will dominate statistical errors.
Therefore, to realise the potential of galaxy clusters as precision probes of cosmology and to maximise the scientific return from future surveys we require a thorough examination of all potential sources of systematics.

Numerical simulations provide a vital resource in this regard: galaxy clusters with exactly known properties.
Cosmological simulations have now matured to the point that many independent groups have simulated either a sufficiently large volume or performed targeted zoom simulations that yield large samples of realistic galaxy clusters, i.e. their properties are broadly matched with observed clusters \citep[e.g.][]{Planelles2013,LeBrun2014,Pike2014,McCarthy2017,Barnes2017a,Barnes2017b,Springel2018,Henden2018,Cui2018,Tremmel2019}.
Lacking both the computational power and a detailed understanding of the physics, simulations model the astrophysical processes known to be important in structure formation in a subgrid manner and calibrate the relevant free numerical parameters on key observational scaling relations, such as the stellar mass-halo mass or gas mass-halo mass relations \citep[e.g.][]{Vogelsberger2014,Schaye2015,Pillepich2018a,Dave2019}.
Therefore, independent models may produce similar global effective behaviours, such as the formation of a quiescent galaxy population or the expulsion of baryons from massive haloes, but they can vary dramatically in the details of how this behaviour is realised.
Although simulations may not capture the full picture of galaxy and structure formation, the plethora of statistically large simulated galaxy cluster samples provide an excellent resource for exploring systematics that have the potential to impact cosmological constraints from future surveys.

However, it is well known that the direct comparison of simulations to observational data is intrinsically flawed, a significant challenge at X-ray wavelengths \citep[e.g.][]{Nagai2007,Khedekar2013}.
The impact of membership, projection effects, multi-temperature structures, and contaminating foregrounds and backgrounds complicate any insights gained from direct comparisons.
To overcome these difficulties, especially at X-ray wavelengths, there have been efforts to create synthetic observations of simulation data \citep[e.g.][]{Rasia2008,Heinz2011,Chluba2012,Biffi2013,ZuHone2014,Torrey2015}.
Synthetic images are produced by projecting the calculated emission spectrum of a source along a chosen line of sight, convolving with the instrumental response and adding any required backgrounds.
However, many synthetic observation codes simply produce the image.
Rather, to quantify any bias or scatter introduced during an observation we require a self-consistent framework that produces images and then derives quantities from them in a manner consistent with observational techniques.
A comparison of synthetic-image-derived quantities to those derived directly from the simulation has the potential to highlight systematics and quantify the scatter introduced by analysis techniques \citep[e.g.][]{Ruppin2019a}.

In this paper we introduce \textsc{Mock-X}, an analysis framework designed to derive observed properties from multi-wavelength synthetic images of numerical simulations.
Designed to explore the systematics that may impact cosmological constraints from future surveys, it yields optical, Compton-$y$ and X-ray images of simulated galaxy clusters and analyses them in a manner consistent with observational techniques.
We will detail how synthetic images are generated from a numerical simulation and how we derive directly observable and reconstructed properties from these images.
Given that one of the most significant systematics is the modelling of baryonic astrophysical processes, the framework is designed to be agnostic to the numerical simulation used as input.
To demonstrate the capabilities of \textsc{Mock-X}, we then present a study of bias introduced by measuring cluster masses from synthetic X-ray observations under the assumption of hydrostatic equilibrium using the IllustrisTNG, BAHAMAS and MACSIS simulations.
We explore the impact of projection on recovered mass and examine how the bias and scatter in estimated mass varies as a function of cluster mass.
In future work, we will explore the redshift evolution of the mass bias (Kannan et al. in prep.), study the multitude of criteria used to define relaxed clusters (Cao et al. in prep.), examine the scatter and covariance of cluster observables (Jorgensen et al. in prep.), and analyze the impact of choosing a cluster centre (Barnes et al. in prep.).

The rest of this paper is structured as follows.
In Section \ref{sec:methods} we provide a brief description of the simulations used throughout this work, outline our sample selection criteria, present how the synthetic images are produced and detail our approach to deriving the thermodynamic profiles and mass estimates of galaxy clusters from synthetic X-ray images.
We then examine the bias introduced by estimating cluster masses assuming that they are in hydrostatic equilibrium in Section \ref{sec:hmb}, comparing to bias found in recent observational studies, exploring the scatter in estimated mass, and examining the impact of cluster projection.
In Section \ref{sec:ssh}, we study how the assumption of spherical symmetry impacts the recovered mass estimate and how selecting relaxed cluster subsets impacts the result.
Finally, in Section \ref{sec:omdb} we examine why the profiles from synthetic X-ray images yield a mass-dependent hydrostatic bias and we present our conclusions in Section \ref{sec:concs}.

\section{Methods}
\label{sec:methods}
In this Section, we outline our numerical approach.
First, we briefly detail the simulations used throughout this work and the halo selection method.
We then define quantities that are computed directly from the simulations, before outlining the methods used for generating synthetic images for each halo.
Finally, we summarize the analysis method for estimating cluster masses from synthetic X-ray images, including the derivation of the thermodynamic profiles from the images, the chosen deprojection method and the mass estimation method.
We note that the chosen X-ray image analysis method is not unique and is one combination of many possible choices \citep[e.g.][]{Ettori2013}.
However, the method outlined below is sufficiently computationally efficient to determine masses for the $\sim14,500$ cluster projections used in this work. 

\subsection{Simulated cluster samples}
One of the most significant systematics is our incomplete knowledge of the physical processes that drive galaxy formation.
Combined with the limited dynamic range afforded by finite computational resources, this necessitates the ``subgrid" approach adopted by all numerical simulations that model galaxy formation on cosmological scales.
A significant step forward in the last decade has been the development of calibrated subgrid galaxy formation models by independent groups.
Calibrated subgrid models adjust their numerical free parameters until they match a limited set of key observational relations, such as the stellar mass function or gas mass-halo mass relation.
Although most models reproduce the overall effects of key astrophysical processes, like galactic winds or the formation of the quiescent galaxy population, the detailed implementation can vary dramatically, such as the feedback channels for stars and active galactic nuclei (AGN).
Therefore, throughout this work, we make use of three numerical simulations to explore the impact of different subgrid models and other associated numerical choices, such as hydrodynamic scheme and numerical resolution.
Below, we briefly describe the subgrid models employed by these simulations but refer the reader to the relevant papers for a more detailed explanation.
Key properties of the simulations used in this work are summarised in Table \ref{tab:simprops}.
We note that the small differences in adopted cosmologies between the simulations have a negligible impact on the results presented in this paper.
We then outline the sample selection method applied to all simulations.

\renewcommand\arraystretch{1.25}
\begin{table}
 \caption{Table summarising key properties of the different numerical simulations used throughout this work. From left to right the columns present the simulation, (target) gas mass, gas softening length, dark matter mass, dark matter softening length and the number of clusters (projections) selected at $z=0.1$, respectively.}
 \centering
 \begin{tabularx}{\columnwidth}{Z{1.31cm} Y{0.96cm} Z{0.3cm} Y{0.96cm} Z{0.3cm} Y{2.25cm}}
 \hline
 Simulation & $m_{\mathrm{gas}}$ & $\epsilon_{\mathrm{gas}}$ & $m_{\mathrm{DM}}$ & $\epsilon_{\mathrm{DM}}$ & $N_{\mathrm{clusters}}$ $(N_{\mathrm{proj}})$ \\
  & $[\mathrm{M}_{\astrosun}]$ & $[\mathrm{kpc}]$ & $[\mathrm{M}_{\astrosun}]$ & $[\mathrm{kpc}]$ & $[M_{200}\geq10^{14}\,\mathrm{M}_{\astrosun}]$ \\
 \hline
 TNG300-l1 & $1.1\times10^{7}$ & $0.37$ & $5.9\times10^{7}$ & $1.48$ & $250$ $(1500)$ \\
 TNG300-l2 & $8.8\times10^{7}$ & $0.74$ & $4.7\times10^{8}$ & $2.95$ & $250$ $(1500)$\\
 TNG300-l3 & $7.0\times10^{8}$ & $1.48$ & $3.8\times10^{9}$ & $5.90$ & $242$ $(1452)$\\
 BAHAMAS & $1.2\times10^{9}$ & $5.96$ & $6.6\times10^{9}$ & $5.96$ & $1994$ $(11964)$\\
 MACSIS & $1.2\times10^{9}$ & $5.96$ & $6.6\times10^{9}$ & $5.96$ & $390$ $(2340)$\\
 \hline
 \end{tabularx}
 \label{tab:simprops}
\end{table}
\renewcommand\arraystretch{1.0}

\subsubsection{IllustrisTNG}
The IllustrisTNG project \citep{Marinacci2018,Naiman2018,Nelson2018,Pillepich2018b,Springel2018} is the successor of the Illustris project \citep{Vogelsberger2014nat,Vogelsberger2014,Genel2014,Sijacki2015}.
The updated galaxy formation model \citep{Pillepich2018a,Weinberger2017} has been shown to reproduce a wide range of observable properties from dwarf galaxies to cluster scales.
In particular, for galaxy clusters, it has been shown to reproduce the metal content of the ICM \citep{Vogelsberger2018} and yield reasonable cool-core fractions at low-redshift \citep{Barnes2018}.
In this work, we make use of all three resolution levels (L1, L2, and L3) of the largest volume simulation - TNG300, a periodic cubic volume with a side length of $302.6\,\mathrm{Mpc}$.
All IllustrisTNG simulations use a cosmological model whose parameters are chosen in accordance with the \citet{Planck2016} constraints: $\Omega_{\mathrm{b}}=0.0486$, $\Omega_{\mathrm{M}}=0.3089$, $\Omega_{\mathrm{\Lambda}}=0.6911$, $\sigma_{8}=0.8159$, $H_{0}=100\,h=67.74\,\mathrm{km}\,\mathrm{s}^{-1}\,\mathrm{Mpc}^{-1}$, and $n_{\mathrm{s}}=0.9667$.

Employing an updated version of the Illustris galaxy formation model \citep{Vogelsberger2013,Torrey2014}, the IllustrisTNG subgrid model includes a new radio mode AGN feedback scheme \citep{Weinberger2017}, a re-calibrated SN wind model and an extended chemical evolution scheme \citep{Pillepich2018a}, magnetic fields \citep{Pakmor2013} and refinements to the numerical scheme that improve its convergence properties \citep{Pakmor2016}.
The magneto-hydrodynamics equations are evolved with the moving-mesh code \textsc{Arepo} \citep{Springel2010}.
The dark matter particles have a mass, $m_{\mathrm{DM}}$, of $5.9\times10^{7}\,\mathrm{M}_{\astrosun}$, $4.7\times10^{8}\,\mathrm{M}_{\astrosun}$ and $3.8\times10^{9}\,\mathrm{M}_{\astrosun}$ for the level 1, 2 and 3 resolutions, respectively.
The collisionless particles, i.e. dark matter and stars, have softening lengths, $\epsilon_{\mathrm{DM}}$, of $1.48\,\mathrm{kpc}$, $2.95\,\mathrm{kpc}$ and $5.90\,\mathrm{kpc}$ for levels 1, 2 and 3 respectively, which is a fixed physical length for $z\leq1$ and comoving for $z>1$.
The gas cells have a target mass, $m_{\mathrm{gas}}$, of $1.1\times10^{7}\,\mathrm{M}_{\astrosun}$, $8.8\times10^{7}\,\mathrm{M}_{\astrosun}$ and $7.0\times10^{8}\,\mathrm{M}_{\astrosun}$, and an adaptive comoving softening length, $\epsilon_{\mathrm{gas}}$, that reaches a minimum of $0.37\,\mathrm{kpc}$, $0.74\,\mathrm{kpc}$ and $1.48\,\mathrm{kpc}$ for levels 1, 2 and 3, respectively.
We note that the subgrid model for IllustrisTNG is calibrated once, for the highest resolution TNG100 volume, and remains fixed across the three different resolution levels, enabling a convergence study with chosen numerical resolution. 

\subsubsection{BAHAMAS}
The BAHAMAS project \citep{McCarthy2017} is a suite of periodic cubic volumes with a side length of $596\,\mathrm{Mpc}$ that vary both their cosmological parameters and the free parameters of the subgrid galaxy formation model.
An evolution of the Cosmo-OWLS project \citep{LeBrun2014}, the project was designed to yield large samples of realistic simulated clusters for cluster cosmology studies.
The galaxy formation model was calibrated to ensure that the baryonic content of galaxy clusters is well matched to observed clusters, and the reference calibration of the model reproduces a wide range of observed scaling relations.
In the work, we make use of the reference \textit{Planck} cosmology run.
This assumes a flat $\Lambda$CDM cosmology constrained by \citet{Planck2014}: $\Omega_{\mathrm{b}}=0.0490$, $\Omega_{\mathrm{M}}=0.3175$, $\Omega_{\mathrm{\Lambda}}=0.6825$, $\sigma_{8}=0.834$, $H_{0}=100\,h=67.11\,\mathrm{km}\,\mathrm{s}^{-1}\,\mathrm{Mpc}^{-1}$, and $n_{\mathrm{s}}=0.9624$.

With its origins in the OWLS \citep{Schaye2010} and GIMIC \citep{Crain2009} projects, BAHAMAS evolves the hydrodynamics equations using traditional SPH.
The galaxy formation model includes radiative cooling \citep{Wiersma2009a}, stochastic, metallicity independent star formation that by construction reproduces the Kennicutt-Schmidt relation \citep{Schaye2008}, mass and metal return due to stellar evolution \citep{Wiersma2009b}, kinetic wind supernovae feedback \citep{DallaVecchia2008}, and the seeding, growth and feedback from supermassive black holes \citep{Springel2005b,Booth2009}.
The initial gas particle mass is $1.21\times10^{9}\,\mathrm{M}_{\astrosun}$ and the dark matter particle mass is $6.63\times10^{9}\,\mathrm{M}_{\astrosun}$.
The gravitational softening length is set to $5.96\,\mathrm{comving~kpc}$ for $z>3$ and to the same value in physical $\mathrm{kpc}$ for $z\leq3$.
The minimum smoothing length of the SPH kernel is set to a tenth of the gravitational softening.

\subsubsection{MACSIS}
The rarity of massive galaxy clusters is such that even a volume of $\sim600^{3}\,\mathrm{Mpc}^{3}$ is too small to contain the largest clusters expected to form in a $\Lambda$CDM cosmology.
The MACSIS project \citep{Barnes2017a} was designed to simulate these missing clusters.
From a dark matter only periodic cube with a side length of $3.2\,\mathrm{Gpc}$, all haloes identified by a standard \textit{Friend-of-Friends} (FoF) percolation algorithm whose mass $M_{\mathrm{FoF}}>10^{15}\,\mathrm{M}_{\astrosun}$ were grouped in logarithmically spaced bins with a width $\Delta\log_{10}(M_{\mathrm{FoF}})=0.2$.
All haloes in a bin were selected if the number of objects in the bin was less than $100$, otherwise, $100$ haloes were selected from the bin at random.
The end result of this process is a sample of $390$ massive galaxy clusters.

Each cluster in the sample was resimulated at increased resolution via the zoomed simulation technique \citep{Katz1993,Tormen1997}, with the high-resolution region free of contaminating tidal particles out to $5\,r_{500}$\footnote{The radius $r_{500}$ denotes the radius of sphere that encloses a mass $M_{500}$ and has a mean density equal to $500$ times the critical density of the Universe.}.
The BAHAMAS galaxy formation model was used for the full physics resimulations and the mass and spatial resolution was chosen to be identical to the original BAHAMAS volume.
The cosmology of the MACSIS simulations was not altered from the parent dark matter volume, which assumes a marginally different flat $\Lambda$CDM cosmology constrained by \citet{Planck2014}: $\Omega_{\mathrm{b}}=0.04825$, $\Omega_{\mathrm{M}}=0.307$, $\Omega_{\mathrm{\Lambda}}=0.693$, $\sigma_{8}=0.8288$, $H_{0}=100\,h=67.77\,\mathrm{km}\,\mathrm{s}^{-1}\,\mathrm{Mpc}^{-1}$, and $n_{\mathrm{s}}=0.9611$.
The combination of MACSIS and BAHAMAS spans the complete mass range of galaxy clusters expected to form in a $\Lambda$CDM cosmology.
We note that the MACSIS sample suffers from selection effects, due to haloes being selected by \textit{Friends-of-Friends} mass and the analysis below using spherical overdensity masses.
We highlight where this potentially impacts the results presented.

\subsubsection{Sample selection}
All simulations used in this work identified haloes via a standard \textit{Friend-of-Friends} algorithm run on the dark matter particles, with a typical value of the linking length in units of the mean interparticle separation $(b = 0.2)$. 
Baryonic particles/cells are attached to haloes by locating their nearest dark matter particle.
Self-bound structures were then identified by via \textsc{subfind} \citep{Springel2001,Dolag2009}, with the most massive subhalo in each FoF group defined as the central object, and the remaining self-bound structures being defined as subhaloes.
For all simulations, clusters were selected from the snapshot closest to $z=0.1$ via the mass cut $M_{500}\geq10^{14}\,\mathrm{M}_{\astrosun}$.
With this threshold, the simulated cluster samples contain $250$, $250$, $242$, $1994$, and $390$ clusters for TNG300-1, TNG300-2, TNG300-3, BAHAMAS and MACSIS, respectively.
Throughout this work, we define the cluster centre as the potential minimum, which is chosen to be the most bound particle identified by the \textsc{subfind} algorithm.

\subsubsection{Properties derived from the simulation}
\label{sec:simprops}
In this work, we use or compare to some cluster properties derived directly from the simulations.
Three-dimensional cluster centric radial density and mass-weighted temperature profiles were computed in the range $0-1.5\,r_{\mathrm{500,sim}}$ by binning the gas cells/particles in $50$ linearly spaced radial bins, where the subscript $\mathrm{sim}$ denotes that the value was produced by the \textsc{subfind} algorithm.
The mass-weighted temperature is defined as
\begin{equation}
 T_{\mathrm{mw}} = \frac{\sum_{k} m_{k}T_{k}}{\sum_{k} m_{k}}\:,
\end{equation}
where $m_{k}$ is the mass of the $k$th cell/particle, $T$ is the temperature and the sum runs over all particles in a given radial bin.
Additionally, to ensure that the fitting of APEC templates to synthetic X-ray spectra is computationally efficient we compute $2$D density, temperature and metallicity cluster centric radial profiles by projecting the cells/particles along the axis of interest and then binning them into $50$ linearly spaced radial bins over $0-1.5\,r_{\mathrm{500,sim}}$.

We compute a theoretical criterion for defining a cluster as relaxed: the energy ratio $E_{\mathrm{rat}}$.
The energy ratio \citep{Barnes2017b} is the kinetic energy of the gas within the cluster divided by its thermal energy. 
When clusters collapse or undergo significant mergers there will be substantial energy stored in the random kinetic motions of the gas.
Over time, this energy is converted to thermal energy via weak shocks \citep[e.g.][]{Kunz2011} and potentially turbulent cascades \citep{Zhuravleva2014} as the cluster relaxes.
Therefore, relaxed clusters should have a lower ratio of kinetic to thermal energy relative to disturbed objects.
We measure this ratio within $r_{\mathrm{500,sim}}$ by removing the bulk motion of the cluster and then computing the ratio of the sum of kinetic energy to the sum of thermal energy for all gas cells/particles that lie within the $3$D aperture
\begin{equation} \label{eq:Erat}
E_{\rm rat}=E_{\mathrm{kin,500}}\,/\,E_{\mathrm{thm,500}}\:.
\end{equation}
In this work, we define as cluster as relaxed if it $E_{\mathrm{ratio}}<0.1$.
Unless otherwise stated, all other cluster properties used in this work are derived from the synthetic images.

\subsection{Synthetic image generation}
For every cluster in the simulation samples, we generate synthetic images that form the basis of our analysis.
The centre of every image is taken to be the projection of the potential minimum.
We note that how to define the centre of a cluster and how derived quantities depend on this choice \cite[e.g.][]{Ruppin2019b} is still an open question, but we leave this issue to future work.

To investigate the impact of projection effects we create $6$ projections of every halo. 
We create $3$ projections along $x$, $y$ and $z$. In addition, we characterize the shape of every cluster via the mass distribution tensor $\mathcal{M}$, or equivalently the inertia tensor $\mathcal{I}$ \citep[e.g.][]{Bett2007}.
Modelling the cluster as a uniform ellipsoid, the mass distribution tensor is a square matrix with components:
\begin{equation}
 \mathcal{M}_{ij}=\sum_{k=1}^{N_{200}}m_{k}\,r_{k,i}\,r_{k,j}\:,
\end{equation}
where $m_{k}$ is the mass of the $k$th cell/particle, $r_{k,i}$ is the $i$th component of the position vector $\textbf{r}_{k}$ in cluster centric coordinates and the sum is over the number of particles, $N_{200}$, within $r_{\mathrm{200,sim}}$.
The square roots of the eigenvalues $A$, $B$ and $C$ of matrix $\mathcal{M}$ are the lengths of the semiprincipal axes, with the convention that $A\geq B\geq C$.
Via the mass tensor, we rotate the particle/cell distribution for every cluster and produce a further $3$ projections along the semi-principal axes $A$, $B$ and $C$.
Additionally, we parameterize how spherical every cluster is by computing its sphericity $s=(C/A)^{\sqrt{3}}$, where a perfectly spherical cluster would have $s=1$.

\begin{figure*}
 \centering
 \includegraphics[width=\textwidth,keepaspectratio=True]{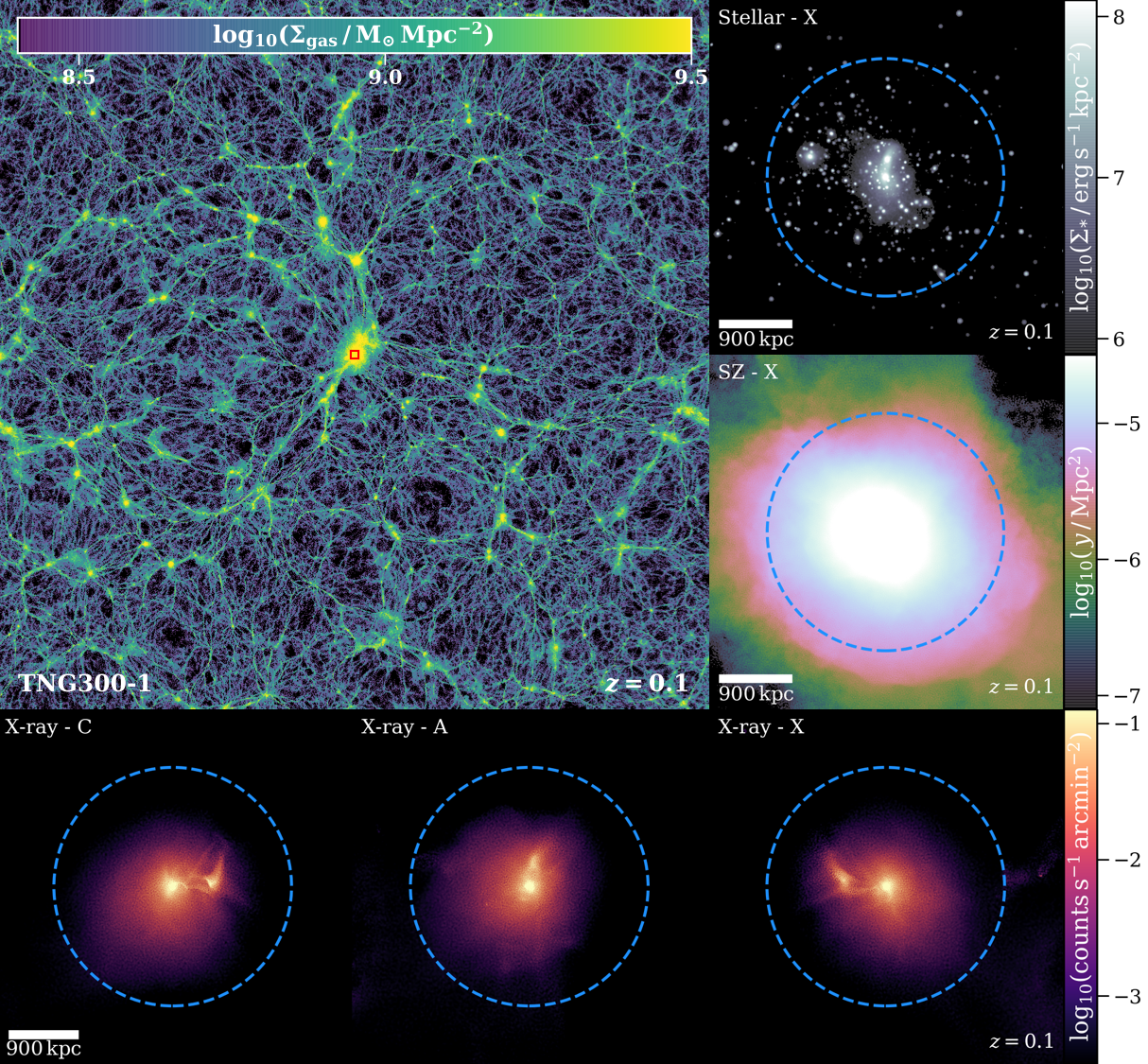}
 \caption{\textsc{Mock-X} schematic from the IllustrisTNG $300^{3}\,\mathrm{Mpc}^{3}$ volume. \textit{Top left}: Gas density slice of $1\,\mathrm{Mpc}$ width through the simulation volume centred on the most massive cluster at $z=0.1$. The red square denotes the $3\,r_{\mathrm{500,sim}}$ square field-of-view used for synthetic image generation. \textit{Top right}: Synthetic optical image of the cluster showing the bolometric luminosity of the star particles. \textit{Middle right}: Synthetic Compton-$y$ image for a NIKA2-\textit{like} facility. \textit{Bottom row}: Synthetic X-ray images for a Chandra-\textit{like} telescope projected along the shortest axis ($C$, left), longest axis ($A$, middle) and the $x$ axis (right) of the simulation volume. The dashed blue line in the synthetic image panels denotes $r_{\mathrm{500,sim}}$.}
 \label{fig:Sch}
\end{figure*}

A schematic of our approach is shown in Fig. \ref{fig:Sch} for the IllustrisTNG level 1 simulation, with the left panel showing a $1\,\mathrm{Mpc}$ depth slice through the gas column density at $z=0.1$.
In the centre of the image is the most massive cluster in the simulation volume, with the red square denoting $3\,r_{\mathrm{500,sim}}$.
The right-hand column of three images shows a synthetic bolometric optical luminosity image, the Compton-$y$ image and a smoothed soft-band $(0.5-2.0\,\mathrm{keV})$ X-ray photon counts image of this cluster projected down the $z$ axis.
The bottom left and bottom middle panels show the X-ray counts projected down the longest $(A)$ and shortest $(C)$ axes through the cluster.

We note that in this work we assume perfect signal to noise, i.e. the synthetic images have no background.
For the X-ray images explored in this work the lack of background produces a minor reduction in the scatter but does not bias the results \citep[e.g.][]{Rasia2008}.
Therefore, we leave the description of how we generate background noise to future work where it is relevant.

\subsubsection{Optical}
Following previous work \citep[e.g.][]{Torrey2015}, we generate optical images by treating every star particle as a stellar population with a single metallicity and age given by the simulation and assuming a \citet{Chabrier2003} initial mass function.
For each star particle its bolometric luminosity and its luminosity in the Sloan Digital Sky Survey $u, g, r, i, z$ bands is computed.
In principle, the light from every star particle should be attenuated by intervening dust and gas.
To accurately model this obscuration we require detailed knowledge of the gas and dust distributions on sub-parsec scales and should perform detailed radiative transfer calculations.
However, all of the simulations used in this work lack the spatial resolution and the self-consistent dust physics to perform such calculations.
Instead, we estimate the impact of dust obscuration by calculating the column density of dust along the line of sight from every star \citep[e.g.][]{Hopkins2005,Wuyts2009}.
We note that, given the limited spatial resolution of the simulations and the treatment of the interstellar medium as an effective single-phase medium, this is still a relatively naive calculation of the optical depths to star particles.

For all star particles with the FoF group, the stellar light is then projected along the relevant axis and smoothed to create maps in all six bands.
The smoothing length of every star particle of interest was computed using a k-d tree.
Throughout this project we determine a star's smoothing length by computing the distance to the $32^{\mathrm{nd}}$ nearest gas neighbour.
Potentially, the choice of smoothing length can subtly alter the shapes of galaxies.
However, for the cluster scale maps generated here, we find that the choice smoothing length has a negligible effect.
For our fiducial maps, we assume an SDSS-\textit{like} angular pixel resolution of $0.24\,\mathrm{arcsec}$ and use a square field of view of length $3r_{500}$.

\subsubsection{Compton-$y$}
The free electrons in the hot plasma of the ICM interact with the photons of the cosmic microwave background (CMB) via inverse Compton scattering, shifting the energy distribution of the CMB from its black-body spectrum and producing the Sunyaev-Zel'dovich (SZ) effect.
In the \textsc{Mock-X} framework we focus on the non-relativistic, thermal SZ effect.
Proposed in $1970$ \citep{SunyaevZeldovich1972}, the first cluster was discovered via the SZ effect was presented in \citet{Staniszewski2009} and it is now routinely used to build large catalogues of galaxy clusters \citep[e.g.][]{Reichardt2013,Hasselfield2013,PlanckSZ2016,Bleem2019}.
An advantage of SZ selected cluster samples is that they are relatively unbiased to the presence of a cool-core \citep{Lin2015}.
To compute realistic images of the SZ effect, in principle one should compute the response of every gas cell/particle at several frequencies and extract the Compton-$y$ signal using a match filtering approach.
However, for this initial set of projects, we only desire the integrated Compton-$y$ signal and centre derived from it.
Therefore, we directly compute the expected Compton-$y$ signal expected for every gas cell/particle.
The projected Compton-$y$ signal along a given line of sight, $l$, is proportional to the projected electron pressure
\begin{equation}
 y = \frac{\sigma_{\mathrm{T}}}{m_{\mathrm{e}}c^2}\int P_{\mathrm{e}}(l)dl\:,
\end{equation}
where $\sigma_{\mathrm{T}}$ is the Thompson scattering cross-section, $m_{\mathrm{e}}$ is the mass of an electron, $c$ is the speed of light, $P_{\mathrm{e}}=n_{\mathrm{e}}k_{\mathrm{B}}T$ is the electron pressure, $n_{\mathrm{e}}$ is the electron number density, $k_{\mathrm{B}}$ is the Boltzmann constant and $T$ is the temperature.
The Compton-$y$ parameter is computed for all gas cells/particles within the FoF group and then projected down the relevant axis to yield Compton-$y$ maps.
We create square maps with a physical side length of $3\,r_{500}$ for all clusters with a resolution of $11\,\mathrm{arcsec}$, similar to the spatial resolution of current SZ facilities like NIKA2 \citep{Adam2018} and MUSTANG-2 \citep{Dicker2014}.

\subsubsection{X-ray}
The impact of multi-temperature structure, gas clumping and projection effects are particularly relevant for X-ray observations and synthetic observations are a vital tool in the faithful comparison of simulations to observations.
Our method for generating synthetic X-ray observations mirrors many previous works in this area \citep[e.g.][Pop et al. in prep.]{Gardini2004,Nagai2007,Rasia2008,Heinz2009,Biffi2012,ZuHone2014}.
However, we have optimised it for the large sample sizes used in this work and the high resolution of some clusters, where a single cluster can contain $>10^{7}$ resolution elements within $r_{500}$.

\begin{figure*}
 \centering
 \includegraphics[width=\textwidth,keepaspectratio=True]{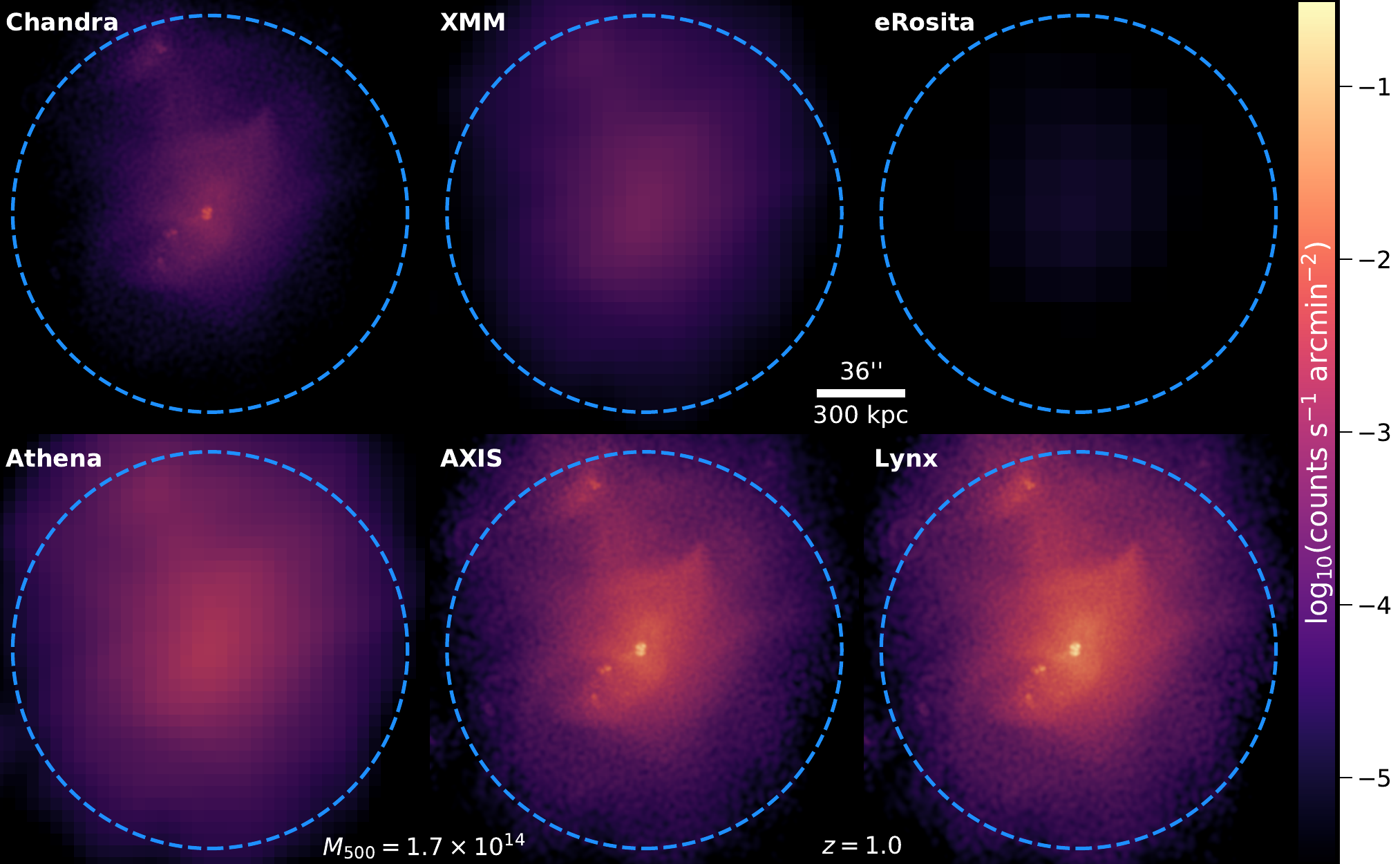}
 \caption{Synthetic X-ray images of a cluster in the TNG300 level 1 volume at $z=1$ for facilities like those currently available and planned. We demonstrate the expected images from \textit{Chandra} (top left), \textit{XXM} (top middle), \textit{eRosita} (top right), \textit{Athena} (bottom left), \textit{AXIS} (bottom middle) and \textit{Lynx} (bottom right) \textit{like} instruments. The blue circle denotes $r_{\mathrm{500,sim}}$. We note the chosen cluster is below the expected detection threshold of \textit{eRosita} at this redshift, with $7$ total photons predicted within $r_{\mathrm{500,sim}}$ for an exposure of $100\,\mathrm{ksec}$. The power of future facilities is demonstrated by the significant increase in photons collected at larger radii.}
 \label{fig:Inst}
\end{figure*}

Similar to previous work \citep[e.g.][]{LeBrun2014,Barnes2017a}, we begin by generating an X-ray spectrum for every gas particle/cell within the FoF group via a lookup table of spectral templates.
We generate the table using the Astrophysical Plasma Emission Code \citep[\textsc{apec};][]{Smith2001} via the \textsc{pyatomdb} module with atomic data from \textsc{atomdb} v3.0.9 \citep[last described in][]{Foster2012}.
For the $11$ chemical elements tracked by the simulations we create a spectral table that spans the temperature range $10^{6}-10^{9}\,\mathrm{K}$ with a spacing $\Delta\log_{10}(T/\mathrm{K})=0.02$.
The energy range and resolution of the spectra depend on the desired instrument, for example, the table of an instrument similar to \textit{Chandra} ACIS-I has an energy range of $0.5-10.0\,\mathrm{keV}$ with an energy resolution of $150\,\mathrm{eV}$.
Improving on previous work, the spectra are convolved with the corresponding response matrix and the effective area for the desired energy bins is taken from the instrument's ancillary response file.
We account for galactic absorption using the \textsc{wabs} model \citep{Morrison1983}, assuming a fixed column density of $n_{\mathrm{H}}=2\times10^{20}\,\mathrm{cm}^{-2}$.
Via this method, it is trivial to precompute spectra lookup tables for current, e.g. \textit{Chandra}, \textit{XMM}, \textit{eRosita}, and proposed, e.g. \textit{Athena}, \textit{Lynx}, \textit{AXIS}, facilities and we demonstrate this by computing the expected emission for a $1.7\times10^{14}\,\mathrm{M}_{\astrosun}$ cluster at $z=1$ for current and (potential) future missions in Fig.\ref{fig:Inst}.
We have deliberately selected a cluster below the detection threshold of the \textit{eRosita} mission, predicting it would see $7$ counts within $r_{500}$ for a $100\,\mathrm{ksec}$ exposure, and the power of future missions is clear, with a significant number photons expected out to $r_{500}$ for \textit{Athena}, \textit{AXIS} and \textit{Lynx}.
Unless otherwise stated, throughout the remainder of this work we assume an instrument similar to \textit{Chandra} ACIS-I.

For every particle/cell we then compute a total spectrum using its temperature, density and metal abundance by summing over the chemical elements tracked by the simulations, which all track the same elements.
We note that we discard the final ``element" tracked by TNG that ensures the total mass of the elements is equal to the mass of the gas cell.
If a particle/cell breaches any of the following conditions we exclude it from the analysis: i) its temperature is less than $10^{6}\,\mathrm{K}$, ii) its star formation rate is non-zero (i.e. it is following an enforced equation of state), or iii) its net cooling rate is positive (i.e. it is increasing in temperature).
These criteria typically remove a few per cent of the total cells/particles within a cluster, ensuring that gas that would not be X-ray emitting or is following an enforced equation of state (i.e. eligible to form stars) is excluded from the image generating process.
Substructures are removed at this stage of image production via \textsc{subfind}, with any gas cells/particles bound to any structure other then the central object removed.

The spectrum for every particle is then projected down the relevant axis and smoothed onto a square grid with a physical side length of $3r_{500}$.
In this work, we neglect issues such as chip gaps, the requirement of stitching multiple pointings together, or that the instrument response can vary across the focal plane.
The pixel resolution of the synthetic image is set by the angular resolution of the chosen instrument, which we set to $0.5\,\mathrm{arcsec}$ for a \textit{Chandra}-like instrument.
Each image pixel stores the combined spectrum of all particles smoothed onto it.
This creates a $3$D datacube, where the first two dimensions are the pixel locations on the image map and the third dimension is the X-ray spectrum of the pixel.
We then assume an exposure time of $100\,\mathrm{ksec}$ to generate the expected photons in each pixel, though with perfect signal to noise this step is academic for this study.

The final result of the synthetic image generation process is a set of images for every simulated cluster: $6$ optical images, a Compton-$y$ image and an X-ray image for each of the $6$ chosen projection axes.
For the remainder of this work, we focus on the estimation of cluster masses from synthetic X-ray images and spectra, assuming hydrostatic equilibrium.
We now outline our analysis method for extracting the thermodynamic profiles of clusters from synthetic X-ray images and how they are used to estimate the mass of the cluster.

\subsection{Hydrostatic masses from Synthetic X-ray images}
We now detail how cluster masses are estimated from the synthetic X-ray images.
We begin by noting that the approach presented below is by no means a unique approach to extracting estimated cluster masses from thermodynamic profiles \citep[e.g.][]{Pointecouteau2005,Nulsen2010,Sanders2018}.
We direct the interested reader to \citet{Ettori2013} for a review of advantages and disadvantages of different approaches.
The approach taken in this work is designed to yield stable mass estimates in a manner that is sufficiently computationally efficient to enable the analysis of $>10^{4}$ mock X-ray datacubes.

\subsubsection{Gas density profiles}
When computing the thermodynamic radial profiles we centre on the potential minimum of the cluster, defined as the most bound particle identified by the \textsc{subfind} algorithm. 
We extract the gas density profile of each datacube from its surface brightness map in the energy band $0.7-1.2\,\mathrm{keV}$.
In this band, the emission is relatively insensitive to the gas temperature and proportional to the square of the density.
The pixels are binned into $25$ linearly spaced annuli in the range $0.0-1.5\,r_{500}$.
We then compute the median surface brightness within each annulus.
In agreement with previous work \citep[e.g.][]{Zhuravleva2013}, we find the median, relative to the mean, to be more robust to the presence of accreting material and the inhomogeneities found in the gas distribution at larger cluster centric radii \citep{Nagai2011,Vazza2013,Roncarelli2013}.
We use the same \textsc{apec} lookup table used to generate the datacubes, with the same galactic absorption model and instrument response and effective area, to model the emission in each annulus as a thin plasma.
The emission measure is then related via
\begin{equation}
 Norm = \frac{10^{-14}}{4\mathrm{\pi}\left[d_{\mathrm{A}}(1+z)\right]^{2}}\int_{V}n_{\mathrm{e}}n_{\mathrm{H}}dV\:,
\end{equation}
where $d_{\mathrm{A}}$ is the angular diameter distance, $n_{\mathrm{e}}$ is the electron number density, and $n_{\mathrm{H}}$ is the ion number density, where we assume $n_{\mathrm{e}}=1.17n_{\mathrm{H}}$ \citep{Anders1989}.
The projected, $2$D radial gas density profile is then computed from the derived emission measure.

\subsubsection{Deprojection}
The modelling of the surface brightness profile yields the $2$D emission measure and density profiles that require deprojection.
Under the assumption of spherical symmetry, the emission measure and density profiles can be deprojected by computing the projected volume $V$ of each spherical shell onto each $2$D annulus.
The profiles are deprojected using the $L1$ regularization method, a non-parametric method that is built upon the work of \citet{Croston2006} and \citet{Ameglio2007}.
For a given $2$D emission measure profile $EMP$, the values of the $3$D emissivity profile $\epsilon$ are then given by maximising the likelihood function
\begin{equation}\label{eq:depro}
 -2\log\mathcal{L}=\chi^{2}=\sum(\mathcal{V}\times\epsilon-EMP)^{2}+\lambda\sum\left|\frac{\partial^{2}\log\epsilon}{\partial\log r^{2}}\right|\:,
\end{equation}
where $\mathcal{V}_{i,j}$ is the geometrical matrix volume of the $j$th shell intercepted by the $i$th annulus, $\times$ denotes a matrix product, and the sum is performed over all annuli.
The second derivative of the emissivity is computed numerically as the derivative of $\epsilon(r)$.
The first term on the right-hand side of equation \ref{eq:depro} for observations is typically divided by the uncertainties, however a perfect knowledge of the $2$D emission measure profile is assumed throughout this work.
The second term on the right-hand side is a penalty term introduced to kill spurious small-scale fluctuations in the recovered $3$D profile \citep{Diaz-Rodriguez2017}.
The parameter $\lambda$ determines the degree of regularisation of the recovered profile and setting it zero makes the method equivalent to the onion-peeling technique \citep[introduced in][]{Kriss1983,Ettori2002,Ettori2010}.

We have compared this approach to the multiscale fitting deprojection method presented in \citet{Eckert2016} and \citet{Ghirardini2019}.
In general, we find good agreement between the recovered $3$D gas density profiles when the two methods are applied to synthetic images of clusters with more regular morphologies.
However, we found that for disturbed images the multiscale approach led to significantly more smoothing of features in the density profile relative to the $3$D density profile produced directly from the simulation.
Additionally, the multiscale fitting approach for disturbed clusters was found to be more computationally expensive for disturbed clusters.
Therefore, we adopt the $L1$ approach for deprojection throughout this work.

\subsubsection{Temperature profile extraction}
To extract a temperature profile from a synthetic image, the pixels were first binned into $25$ linearly spaced annuli in the range $0.0-1.5\,r_{500}$.
In each annulus, the pixel spectra were summed to yield a single spectrum for each annulus.
The spectrum was then modelled as a single-temperature plasma with the temperature, emission measure and metallicity free to vary.
We leverage the $2$D mass-weighted temperature, emission measure and metallicity profiles extracted directly from the simulation to provide an initial starting point for the fit, leading to a significant increase in the computational efficiency of our approach.
The solar abundance of \citet{Anders1989} was assumed.
To mimic the observations more closely, each fit assumes an integration time of $100\,\mathrm{ksec}$ to convert the spectrum from $\mathrm{erg}\,\mathrm{s}^{-1}$ to photon count in each energy bin, using the midpoint of the energy as the assumed photon energy.
The fit was then performed in the energy range $0.5-10.0\,\mathrm{keV}$ for those energy bins that contained at least $1$ count, making use of spectra interpolated from the APEC lookup table.

\begin{figure*}
 \centering
 \includegraphics[width=\textwidth,keepaspectratio=True]{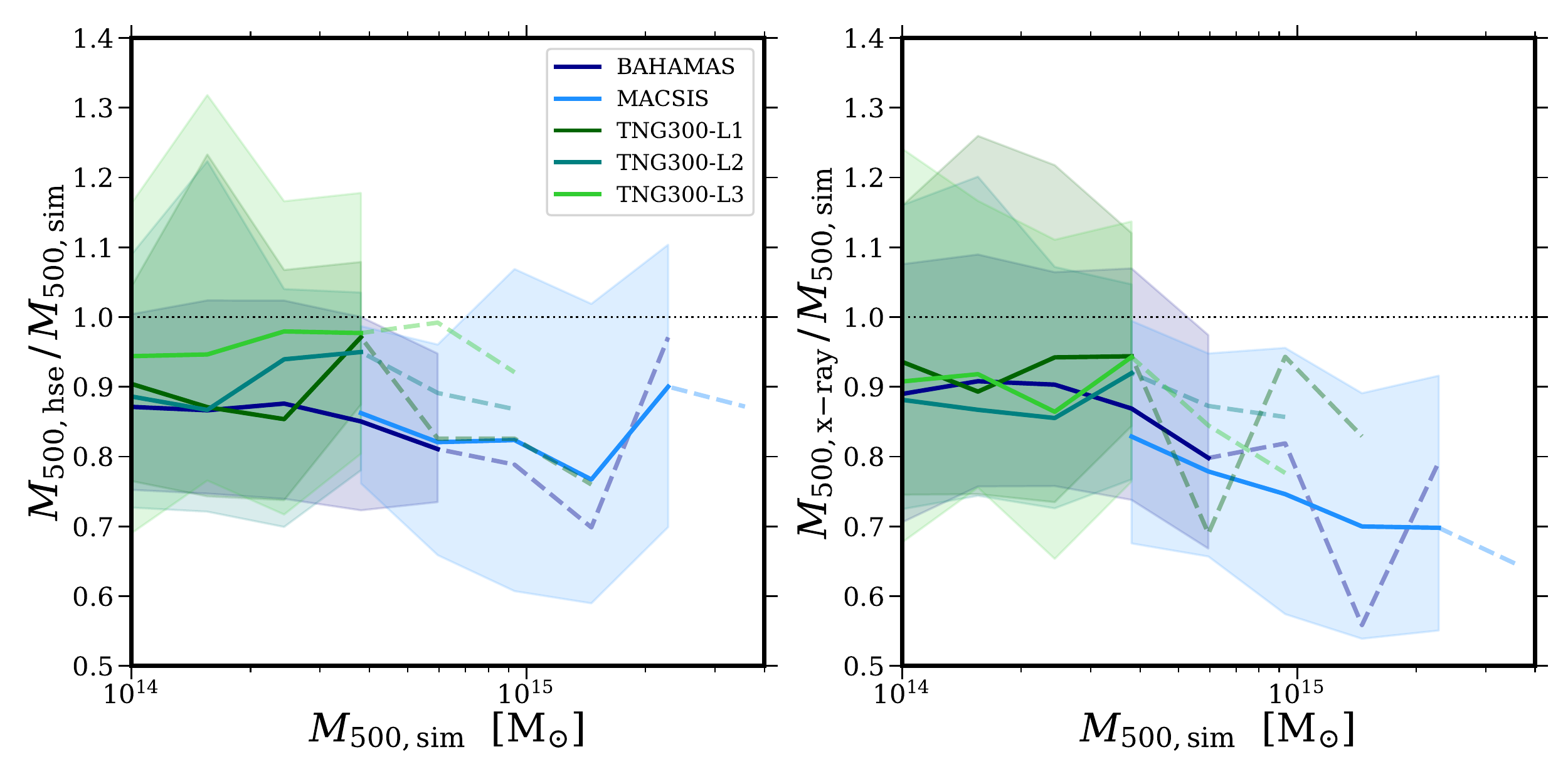}
 \caption{The median hydrostatic mass estimate to true mass $(M_{\mathrm{500,sim}})$ ratio as a function of true mass at $z=0.1$ for estimates derived from simulation (left) and synthetic X-ray image (right) profiles. We plot simulated samples from BAHAMAS (dark blue), MACSIS (light blue), TNG300 level 1 (dark green), level 2 (medium green), and level 3 (light green). The shaded area denotes the $1\sigma$ scatter and the dashed lines denote the median ratio where the number of clusters in a bin of width $\Delta\log_{10}(M_{\mathrm{500,sim}}\,/\,\mathrm{M}_{\astrosun})=0.1$ is less than 10. The hydrostatic bias is relatively independent of true mass for profiles derived directly from the simulation but is clearly mass-dependent for synthetic X-ray profiles.}
 \label{fig:Mbias}
\end{figure*}

The method yields a spectroscopic temperature measurement in each annulus, the combined result of fitting all annuli produces the $2$D temperature profile.
This profile is then deprojected using the projection matrix $\mathcal{V}$ weighted by the emissivity produced when deprojecting the gas density profile.
Following \citet{Mazzotta2004} and \citet{Ghirardini2019}, the $2$D spectroscopic temperature profile is weighted in a spectroscopic-\textit{like} fashion during the deprojection, i.e.
\begin{equation}
 T_{\mathrm{sl}} = \frac{\sum n_{\mathrm{e}}^{2}T^{-1/2}}{\sum n_{\mathrm{e}}^{2}T^{-3/2}}\:.
\end{equation}
This yields a $3$D spectroscopic temperature profile for each synthetic image.
The analysis produces spectroscopic measurements of the thermodynamic profiles for every projection.
Under the assumption that the cluster is in hydrostatic equilibrium, we then compute an estimated mass for each projection using the $3$D profiles derived both directly from the simulation and those computed from the synthetic X-ray images.

\subsection{Hydrostatic mass estimation}
Deriving a cluster's mass assuming spherical symmetry and that it is in hydrostatic equilibrium is a well-established technique.
There have been many observational \citep[e.g.][and references therein]{Miyatake2019} and theoretical \citep[e.g.][]{Nelson2014,Biffi2016} studies that have explored the bias induced by the required assumptions.
Following \citet{Vikhlinin2006}, the $3$D gas density profile is modelled via a modified $\beta$-model profile
\begin{equation}\label{eq:Vrho}
 n_{\mathrm{e}}n_{\mathrm{H}}=n_{\mathrm{0}}^{2}\frac{(r/r_{\mathrm{c}})^{-\alpha}}{(1+r^{2}/r_{\mathrm{c}}^{2})^{3\beta-\alpha/2}}\frac{1}{(1+r^{\gamma}/r_{\mathrm{s}}^{\gamma})^{\xi/\gamma}}\:,
\end{equation}
where the value of $\gamma$ is fixed such that $\gamma=3$ and unphysically sharp density breaks are excluded via the condition $\xi<5$.
The best-fit values obtained are then fed into the analytic derivative of eq. (\ref{eq:Vrho}) to yield a smoothly varying estimate of the density gradient.

Additionally, the $3$D temperature profiles are fit with the general model
\begin{equation}\label{eq:Vtemp}
 T_{\mathrm{3D}}(r)=T_{0}t_{\mathrm{cool}}(r)t(r)\:,
\end{equation}
where $T_{0}$ is a model parameter and
\begin{equation}
 t_{\mathrm{cool}}(r)=\frac{x+T_{\mathrm{min}}T_{0}}{(x+1)}\:,
\end{equation}
where $x=(r/r_{\mathrm{cool}})^{a_{\mathrm{cool}}}$ and
\begin{equation}
 t(r)=\frac{r/r_{\mathrm{t}}}{\left[1+(r/r_{\mathrm{t}})^{b}\right]^{c/b}}\:.
\end{equation}
The best fitting model parameters are again used in the analytic derivative of eq. (\ref{eq:Vtemp}) to provide a smoothing varying estimate of the gradient of the temperature.

The cumulative total mass profile of the projection is then computed assuming hydrostatic equilibrium via
\begin{equation}\label{eq:mest}
 M_{\mathrm{tot}}(<r)=\frac{rk_{\mathrm{B}}T(r)}{G\mu m_{\mathrm{p}}}\left[\frac{d\log\rho_{g}}{d\log r}+\frac{d\log T}{d\log r}\right]\:,
\end{equation}
where $r$ is the $3$D radial cluster centric distance in Mpc, $G$ is the gravitational constant, $\mu=0.59$ is the mean molecular weight and $m_{\mathrm{p}}$ is proton mass.
Given the cumulative mass as a function of cluster centric radial distance, the density profile of a given projection can then be computed.
For a given redshift and cosmology, the density is divided by the critical density and $r_{\mathrm{500,est}}$ is found by interpolating the profile to the radius at which it crosses a value of $500$.
The estimated cluster mass, $M_{\mathrm{500,est}}$, is then given by the summing the cumulative mass profile to the value of $r_{\mathrm{500,est}}$.
Depending on the profiles used for the estimate, we label the recovered mass and radius as follows.
For thermodynamic profiles extracted directly from the simulation, we denote them as $M_{\mathrm{500,hse}}$ and $r_{\mathrm{500,hse}}$, respectively, while for spectroscopic profiles extracted from the synthetic X-ray images we label them as $M_{\mathrm{500,x-ray}}$ and $r_{\mathrm{500,x-ray}}$, respectively.
For each cluster, we compute $6$ mass estimate using the deprojected spectroscopic profiles and $1$ estimate from the mass-weighted profiles.

\subsection{Centroid shift criterion}
\label{sec:csc}
The centroid shift criterion is used to classify clusters as dynamically relaxed and is computed from the surface brightness maps of galaxy clusters.
It has been used extensively in the literature \citep[e.g.][]{Mohr1993,Thomas1998,Poole2006,Kay2007,Maughan2008,Bohringer2010,Rasia2013}.
The criterion measures the deviation of the X-ray centroid as the radius of the chosen aperture reduces.
The centroid shift measures the circular symmetry of the X-ray emission and tests for the presence of significant X-ray emission associated with larger substructures.
We compute the centroid shift via
\begin{equation}
 w = \frac{1}{r_{\mathrm{500,sim}}}\sqrt{\frac{\sum\Delta_{i}-\langle\Delta\rangle}{M-1}}\:,
\end{equation}
where $\Delta$ is the separation of the centroids, the angle brackets denote the average, and we compute the centroids in circular apertures with radii in the range $0.15-1.0\,r_{\mathrm{500,sim}}$ with steps of $0.05\,r_{\mathrm{500,sim}}$.
A cluster is classified as relaxed if $w<0.01$ \citep[e.g.][]{Weissmann2013}.

\section{Hydrostatic mass bias}
\label{sec:hmb}
We begin by exploring the bias induced by estimating cluster masses assuming that they are in hydrostatic equilibrium.
Throughout this work we define the bias as $b=1-M_{\mathrm{500,est}}/M_{\mathrm{500,sim}}$, and we denote the \textsc{subfind} mass, $M_{\mathrm{500,sim}}$, as the ``true" cluster mass.
Using density and temperature profiles extracted from the simulation, we find that for IllustrisTNG there is an excellent agreement in the amplitude of the median bias for the highest resolution runs with $b=0.11\pm0.01$ and $b=0.11\pm0.01$ for levels 1 and 2, respectively.
We find a decrease in the median bias for the lowest resolution, with $b=0.05\pm0.01$ for level 3, but this difference is significantly smaller than the scatter in the population.
The uncertainty on the bias is computed via bootstrap resampling the cluster populations $10,000$ times.
The BAHAMAS sample yields a median bias of $b=0.13\pm0.002$, in good agreement with the IllustrisTNG level 1 result.
Finally, the MACSIS sample yields a median bias of $b=0.15\pm0.003$.
In the left panel of Fig. \ref{fig:Mbias}, we plot the ratio of the estimated mass from simulation derived profiles to true mass as a function of true mass.
All simulations show a significant scatter in the estimated to true mass ratio, with TNG yielding a slightly larger scatter relative to BAHAMAS at fixed mass.
The MACSIS sample yields a larger scatter relative to BAHAMAS, which may be due to the selection function of the MACSIS haloes.
We find no obvious trend of hydrostatic bias with mass when using the thermodynamic profiles derived directly from the simulation.
The choice of numerical resolution, hydrodynamic scheme and subgrid physics reassuringly appears to have minimal impact on the hydrostatic bias recovered, under the assumption that the simulation is being performed at, or near, the numerical resolution level at which the subgrid model was calibrated.

If the spectroscopic density and temperature profiles, estimated from synthetic X-ray projections, are used to compute an estimated mass, we find that the IllustrisTNG samples yield similar results, with a median bias of $b=0.10\pm0.01$, $b=0.12\pm0.01$ and $b=0.08\pm0.01$ for levels 1, 2 and 3, respectively.
The BAHAMAS sample yields a bias of $b=0.11\pm0.003$, consistent with the value from simulation derived profiles.
However, the median bias of the MACSIS sample increases to $0.25\pm0.005$.
In the right panel of Fig. \ref{fig:Mbias}, we plot the ratio mass estimated via spectroscopic profiles to true mass as a function of true mass and we find that hydrostatic bias is now mass-dependent.
For low mass clusters $(M_{\mathrm{500,sim}}<3\times10^{14}\,\mathrm{M}_{\astrosun})$ the bias is consistent between the different samples, with TNG again yielding slightly larger scatter.
However, above this mass the hydrostatic bias begins to increase and for the most massive clusters $(M_{\mathrm{500,sim}}>2\times10^{15}\,\mathrm{M}_{\astrosun})$ the bias is as large as $b=0.3$.
Although lacking the statistics for a rigorous comparison, all of the simulation samples show some evidence that the bias is increasing with cluster mass.
The mass dependence of the hydrostatic mass bias is consistent with previous numerical work that computed estimated cluster masses from spectroscopic profiles \citep{Henson2017,Barnes2017b,Pearce2020}, although these works did not account for the impact of projection.
We now compare our results to those of previous work.

\subsection{Comparison to previous results}
Observationally, there are many approaches to estimating a cluster's mass. 
Relative mass calibration is very common and derives a cluster's mass by equating an observable property, such as X-ray temperature or integrated Compton-$y$ signal, to an observable-mass scaling relation.
However, this method is only possible once a scaling relation is calibrated via absolute mass measurements.
Absolute mass calibration measurements typically require very high-quality observational data.
The member galaxies of a cluster can be used to estimate its mass, either through measuring the richness \citep[e.g.][]{Yee2003,Simet2017}, i.e. the number of galaxies present, or by using them as kinematic tracers of the underlying potential \citep[e.g.][]{Diaferio1997,Zhang2011,Bocquet2015,Sereno2015,Gifford2017}.
However, this approach is fundamentally limited by determining cluster membership \citep[e.g.][]{Old2014} and the fact that theoretical studies have shown galaxies are biased tracers of the underlying potential, agreeing on the size of the bias but not its sign \citep{Munari2013,Armitage2018}.
A study of caustic mass estimates to hydrostatic mass estimates by \citet{Maughan2016} favoured a hydrostatic bias of less than $10$ per cent, though we caution that a recent detailed study of the X-COP clusters found caustic masses underestimate hydrostatic mass significantly \citep{Ettori2019}.

\begin{figure}
 \centering
 \includegraphics[width=\columnwidth,keepaspectratio=True]{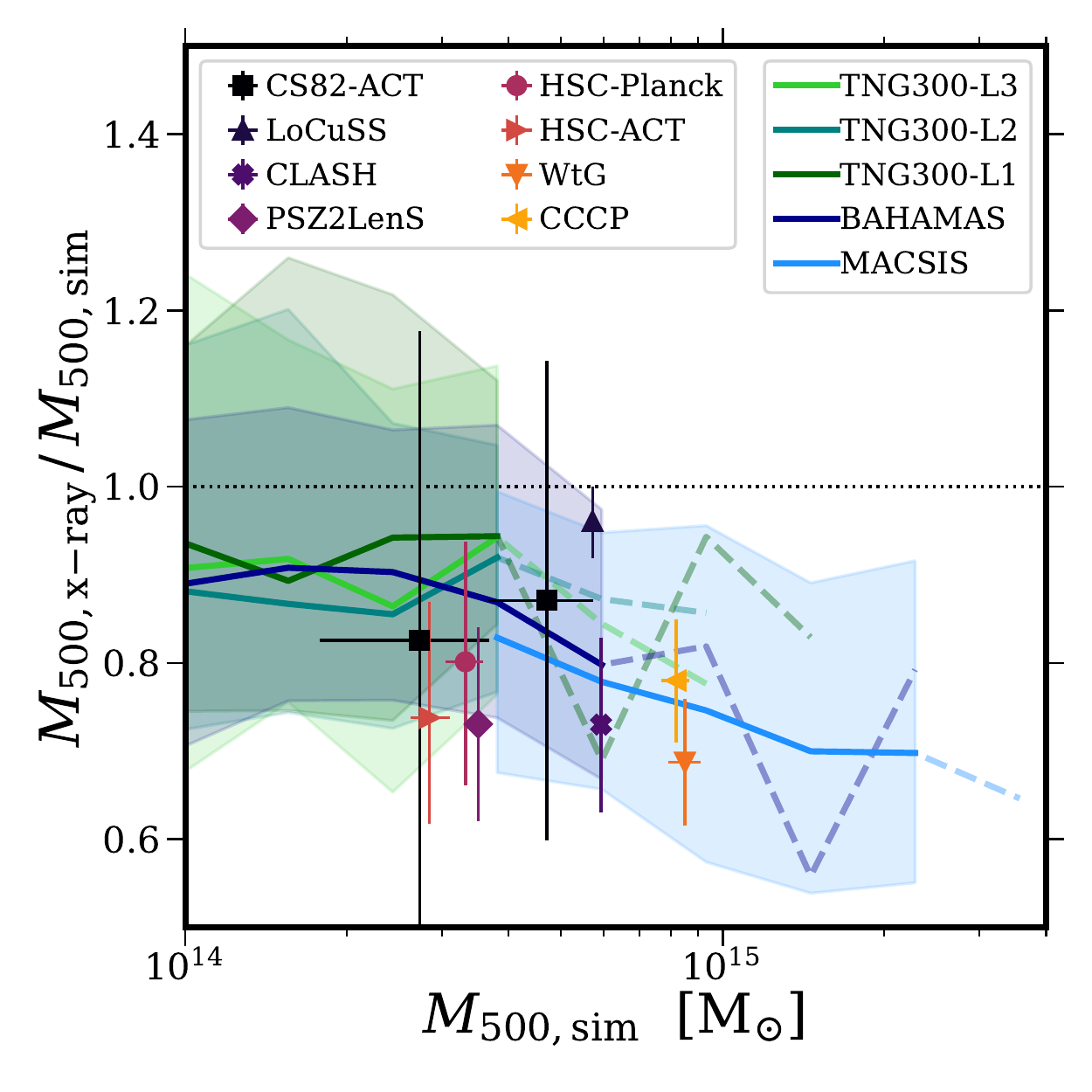}
 \caption{The median hydrostatic mass derived from X-ray images to true mass ratio as a function of true mass at $z=0.1$ against a collection of observed biases. The simulation lines styles are the same as Fig. \ref{fig:Mbias}. The collection of observational points compare hydrostatic mass estimates to weak lensing derived mass estimates and are taken from Weighing the Giants (WtG) \citep{vonderLinden2014}, CCCP \citep{Hoekstra2015}, CS82-ACT \citep{Battaglia2016}, LoCuSS \citep{Smith2016}, CLASH \citep{PennaLima2017}, PSZ2LenS \citep{Sereno2017}, HSC-Planck \citep{Medezinski2018} and HSC-ACT \citep{Miyatake2019}. We find excellent agreement in the magnitude of hydrostatic bias between the simulated and observed samples. Additionally, the sample variance of the observations is well matched to the scatter of the simulated sample.}
 \label{fig:Obs}
\end{figure}

\begin{figure*}
 \centering
 \includegraphics[width=\textwidth,keepaspectratio=True]{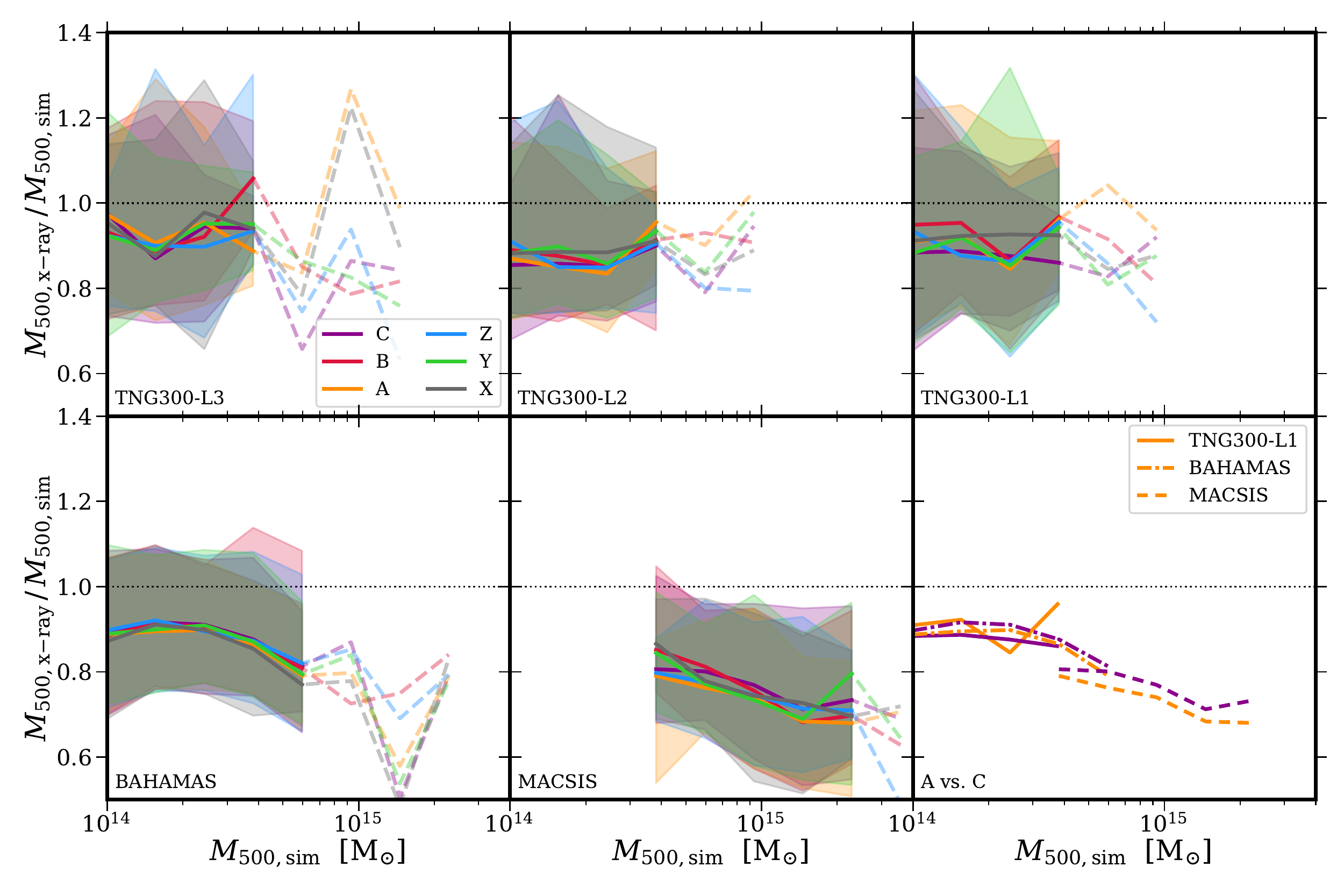}
 \caption{The median X-ray derived hydrostatic mass to true mass ratio as a function of true mass at $z=0.1$, split by simulation and projection axis. For TNG300 levels 3 (top left), 2 (top middle), 1 (top right), BAHAMAS (bottom left) and MACSIS (bottom middle) we plot the median ratio along axes $x$ (grey), $y$ (green), $z$ (blue), $A$ (orange), $B$ (red) and $C$ (purple). The dashed lines denote the median ratio where the number of clusters in a bin of width $\Delta\log_{10}(M_{\mathrm{500,sim}}\,/\,\mathrm{M}_{\astrosun})=0.1$ is less than 10 and the shaded regions denote the $1\sigma$ scatter in the ratio. In the bottom right panel, we compare projections along the longest $(A)$ and shortest $(C)$ axes of the cluster and find minor differences in the median ratio that are significantly smaller than the scatter in any given sample.}
 \label{fig:MBproj}
\end{figure*}

With the advent of deep and wide optical surveys, weak lensing, the statistical distortion of background galaxies due to the intervening mass, has become the absolute cluster mass estimator of choice.
Using the weak lensing cluster mass estimate as a measure of the true cluster mass, it is possible to infer the bias induced by measuring a cluster's mass assuming it is in hydrostatic equilibrium.
Though we caution that theoretical studies have shown that weak lensing mass estimates themselves typically underestimate cluster masses by $5-10$ per cent \citep[e.g.][]{Becker2011,Bahe2012,Rasia2014,Henson2017}.

There is mild disagreement in the magnitude of the hydrostatic bias between observational studies.
Some have found that the bias is less than $10$ per cent, with \citet{Applegate2014}, \citet{Israel2014} and \citet{Smith2016} finding values of $b=0.04$, $b=0.08$ and $b=0.05$, respectively.
However, others have found biases greater than $20$ per cent, with studies by \citet{Mahdavi2008}, \citet{Mahdavi2013}, \citet{vonderLinden2014}, \citet{Hoekstra2015}, \citet{Simet2015}, \citet{Battaglia2016}, \citet{PennaLima2017}, \citet{Sereno2017}, \citet{Medezinski2018} and \citet{Miyatake2019} all finding significantly larger hydrostatic bias, on the order of $20-30$ per cent.
A potential explanation for part of the differing bias estimates may be attributed to the varying overdensities at which the masses are estimated and the different mass and redshift distributions of the various samples.

In Fig. \ref{fig:Obs} we compare our hydrostatic bias measurement from spectroscopically measured thermodynamic profiles as a function of true cluster mass against a collection of observational results taken from \citet{Miyatake2019}, many of which were also presented in \citet{Medezinski2018}.
We find excellent agreement between the observed hydrostatic bias measurements and those recovered from the synthetic images of simulated samples.
Additionally, we note that the sample variance of the observational results is well matched to the scatter in the simulated results.
We note that scatter in the simulated ratios will increase slightly if lensing masses were used, rather then the true mass, due to the intrinsic scatter in the mass estimate.
However, we delay a detailed study of lensing mass estimates to a future study.
The observed amplitude of the hydrostatic bias is reproduced by the simulations and the presence of a mass-dependent mass bias is certainly not ruled out by the observations, with \citet{Medezinski2018} stating that the observations may suggest a mass-dependent bias.

Our results are also consistent with previous numerical studies.
Those that derive a hydrostatic mass estimates directly from simulation data, typically mass-weighted profiles, find they are biased low relative to true masses by $10-20$ per cent \citep[e.g.][]{Lau2009,Kay2012,Rasia2014,Nelson2014,Biffi2016,Angelinelli2019,Ansarifard2019}.
Additionally, for those studies that compute a cluster's thermodynamic profiles from synthetic X-ray data also find good agreement, biased low at the level of $10-20$ per cent for low mass ($M_{\mathrm{500,sim}}<8\times10^{14}\,\mathrm{M}_{\astrosun}$) clusters \citep[e.g.][]{Nagai2007,Rasia2012,LeBrun2014} with a bias that increases with increasing mass \citep{Henson2017,Pearce2020}.
We note that \citet{Biffi2016} also compute the bias using a spectroscopic-like weighted profiles \citep{Mazzotta2004} and find that the hydrostatic bias in their relatively massive $(M_{\mathrm{200,sim}}>8\times10^{14}\,h^{-1}\,\mathrm{M}_{\astrosun})$ sample increased.
Given the large scatter in estimated mass, the small sample size of many of these studies, typically $<50$, is a limiting factor and the synthetic X-ray approach is typically limited to the direct computation of $3$D profiles, ignoring issues such as projection effects and the presence of clumpy gas.

\subsection{The impact of projection}
All clusters to varying degrees are triaxial in nature and determining the projection angle is very difficult \citep[e.g.][]{Morandi2012,Sereno2017}, although there have been interesting recent works relating the orientation of the brightest cluster galaxy to the cluster triaxiality \citep[e.g.][]{Wittman2019,Herbonnet2019}.
This is potentially an issue for hydrostatic mass estimates due to the assumption that the cluster is spherically symmetric.
The triaxial nature of clusters can be removed from analyses by stacking many clusters into a single profile.
However, in numerical simulations, we know the orientation of the cluster.
Therefore, we now examine the impact of projection on the estimated cluster masses by splitting the synthetic images for the different simulations into their projection axes.
To assess the median hydrostatic bias of randomly orientated haloes we use the Cartesian projection axes $(x,~y,~z)$ as samples of randomly projected clusters, making the fair assumption that a cluster's triaxiality should be independent of its orientation in the simulation volume.
The projections along the eigenvectors of the mass tensor $(A,~B,~C)$ enable an examination of the impact of cluster triaxiality on the estimated mass.

In Fig. \ref{fig:MBproj} we plot the ratio of the estimated mass from synthetic X-ray observations to true mass as a function of true mass for the different samples.
Although somewhat noisy, we find that the median of the ratio of estimated mass to true mass projected along either the $A$ or $C$ axes is consistent with values recovered from projecting along the $x,~y,~z$ axes for all simulations.
Additionally, the scatter in the ratio is also consistent for all the chosen projection axes.
For clarity, in the bottom right panel of Fig. \ref{fig:MBproj} we plot the median ratios for the BAHAMAS, MACSIS and IllustrisTNG level 1 simulations computed for synthetic images projected along the $A$ and $C$ axes.
The average change in the bias from projecting along $C$ rather than $A$ is $\Delta b=-0.01$, $0.02$ and $0.03$ for TNG100-L1, BAHAMAS and MACSIS, respectively.
However, the difference in hydrostatic bias for any of the simulated samples is significantly smaller than the scatter in the population of the cluster sample.
Therefore, we conclude that for a sufficiently large sample of clusters the projection angle has minimal impact on the hydrostatic mass estimate.
On a cluster-by-cluster basis, the scatter in estimated mass will dominate over the projection angle.

\subsection{Estimated mass scatter}
To study the scatter in estimated mass, we first compute a kernel-weighted moving average of the mass estimated from synthetic X-ray observations as a function of true mass for all simulated cluster samples.
For the $i$th cluster of mass $M_{i}$, we average over the neighbouring clusters using of a Gaussian weight function of the form
\begin{equation}
 w_{k} = \frac{1}{\sqrt{2\pi}\tau} \exp\left[-\frac{\mu_k^2}{2 \tau ^2}\right]\:,
\end{equation}
where $\mu_{k}=\log(M_{k}\,/\,M_{i})$ for the $k$th neighbour.
The parameter $\tau$ is a hyperparameter that specifies the width of the kernel, with a small value leading to a noisy moving average and a large value over-smoothing.
After testing, we use a value of $\tau=0.1$ to compute the moving average.
We compute the moving average in log-log space.

We then compute the scatter about the moving average for a set of $17$ points in log mass that are linearly spaced in the range $14.0<\log_{10}(M\,/\,\mathrm{M}_{\astrosun})<15.7$ with $\Delta\log(M)=0.1$ via   
\begin{equation}
 \sigma = \sqrt{\frac{1}{N-2}\sum_{k}^{N}\left[\log_{10}(Y_{k})-\log_{10}(Y_{\mathrm{MA}}(M_{k}))\right]^{2}}\:,
\end{equation}
where the sum runs over the $N$ clusters within a top-hat window function of width $\Delta\log_{10}(M)=0.2$, $Y_{k}$ is the estimated mass for the $k$th cluster and $Y_{\mathrm{MA}}$ is the moving average of the estimated mass at true mass $M_{k}\equiv M_{k,\mathrm{500,sim}}$ of the cluster. We compute the uncertainty on the estimated scatter by bootstrap resampling the samples $10,000$ times. 

\begin{figure}
 \centering
 \includegraphics[width=\columnwidth,keepaspectratio=True]{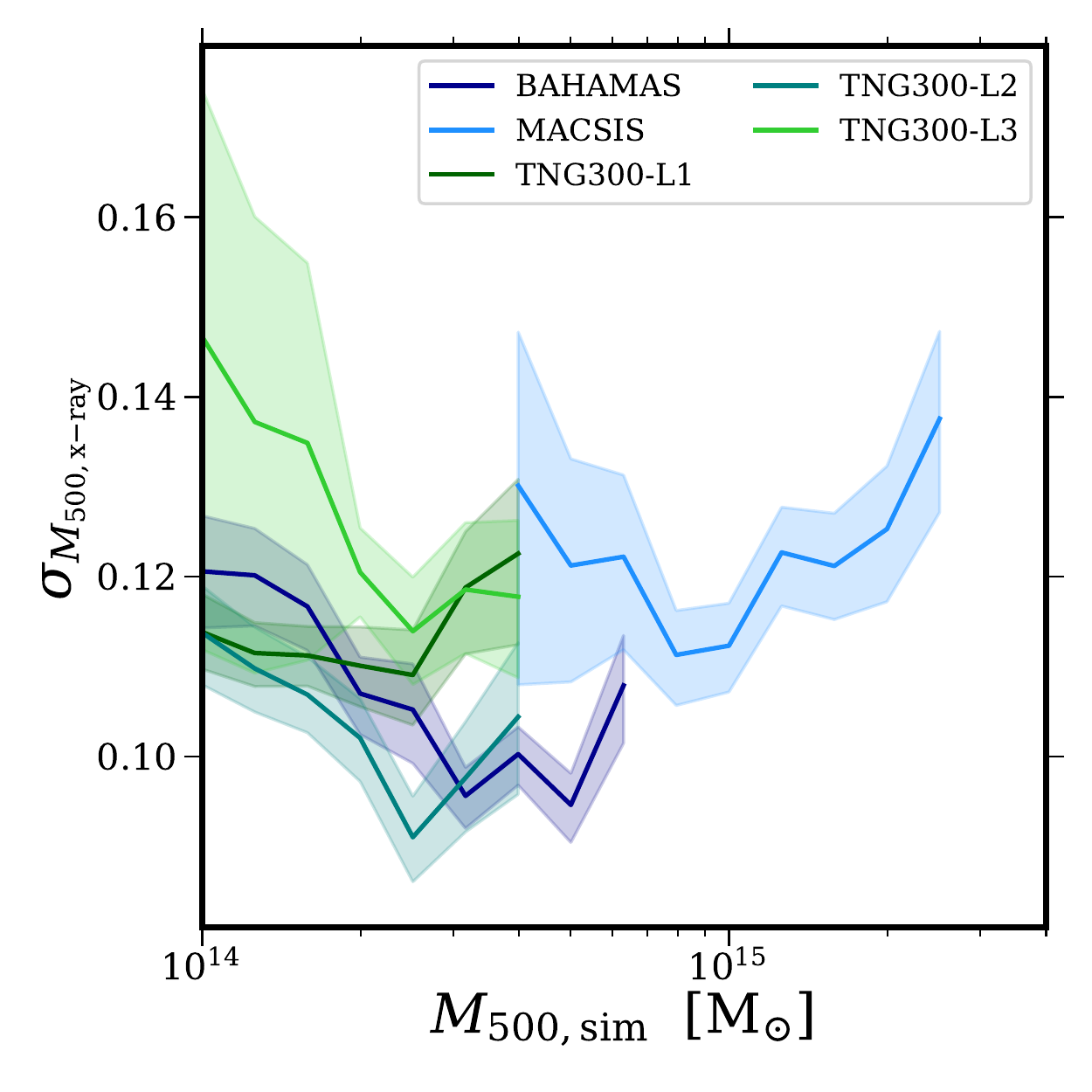}
 \caption{Hydrostatic mass estimate scatter as a function of true mass at $z=0.1$ for the simulation samples. Line styles are the same as Fig. \ref{fig:Mbias}. The scatter does not change significantly with mass, but there is evidence of a minor trend of decreasing scatter with increasing mass before the sample statistics become small and the scatter increases. Selection effects are likely the cause of the increased amplitude of the scatter at fixed mass for the MACSIS sample relative to the BAHAMAS sample.}
 \label{fig:Mscatter}
\end{figure}

\begin{figure*}
 \centering
 \includegraphics[width=\textwidth,keepaspectratio=True]{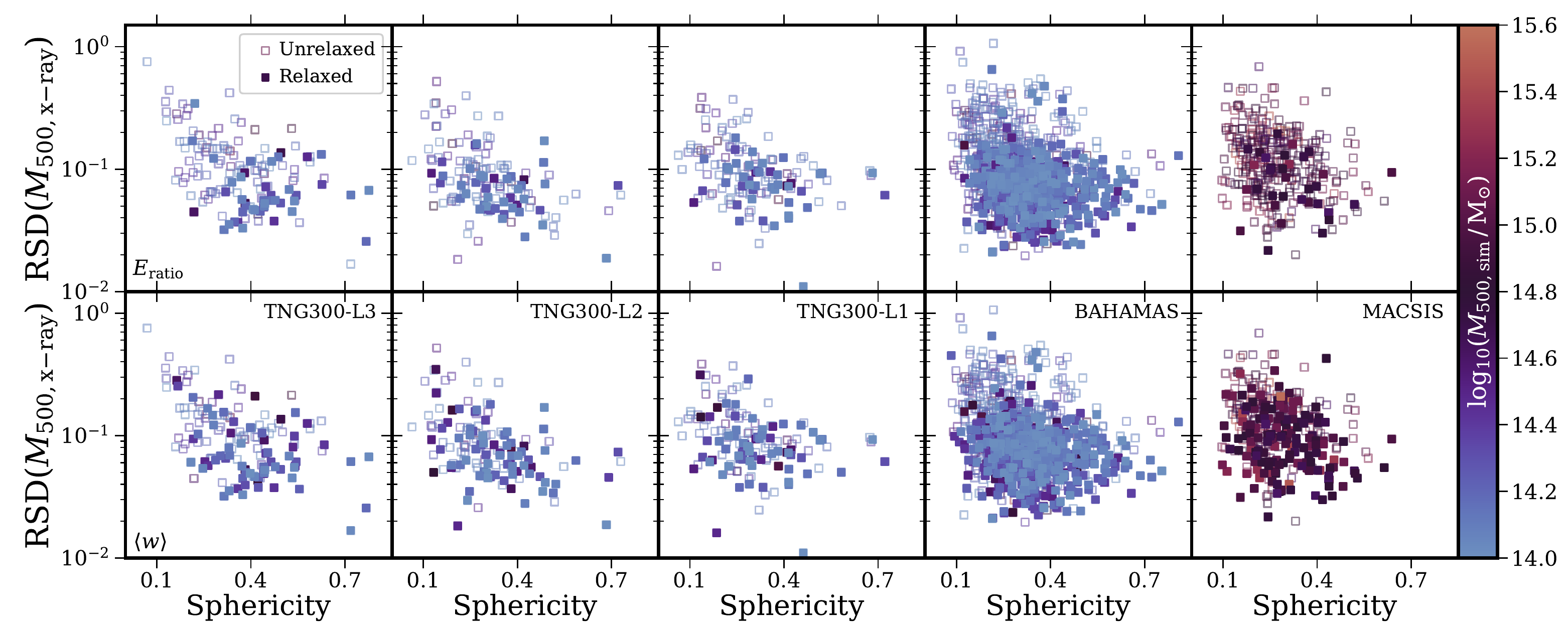}
 \caption{Relative standard deviation (RSD) of the estimated mass between cluster projections as a function of cluster sphericity at $z=0.1$. From left to right we plot the TNG300 level 3, 2, 1, BAHAMAS and MACSIS samples, respectively. The points are coloured by the log of the true mass. The filled (open) points are classified as relaxed (unrelaxed) by the energy ratio (top) and centroid shift (bottom) criteria. We find the expected result that the RSD in the estimated mass reduces as the cluster becomes more spherical. Overall, the relaxation criteria remove those clusters with the largest estimated mass variations, but they do not select more spherical haloes on average nor do they classify the same subset of clusters as relaxed.}
 \label{fig:shape-mass}
\end{figure*}

In Fig. \ref{fig:Mscatter}, we plot the scatter in estimated mass as a function of true mass for all simulated samples.
For IllustrisTNG, we find that the scatter in estimated mass reduces with increasing cluster mass, independent of numerical resolution, but increases in the final mass as the sample size reduces.
The amplitude of the scatter appears to increase with decreasing numerical resolution.
At $M_{\mathrm{500,sim}}=10^{14}\,\mathrm{M}_{\astrosun}$, there is small increase in scatter from level 1 $(\sigma=0.112)$ to level 2 $(\sigma=0.114)$, but a significantly larger increase in scatter for IllustrisTNG level 3 $(\sigma=0.147)$.
The normalization of the estimated mass scatter is slightly larger for the BAHAMAS sample relative to IllustrisTNG level 1, with $\sigma=0.120$ at $M_{\mathrm{500,sim}}=10^{14}\,\mathrm{M}_{\astrosun}$.
However, we find the same trend of decreasing scatter with increasing cluster mass, before increasing again as the sample statistics become small.
The MACSIS sample has greater scatter at fixed mass relative to the BAHAMAS sample, however, this is likely due to the way the sample was selected.
These lower mass MACSIS clusters have a significantly larger FoF mass relative to $M_{\mathrm{500,sim}}$, as they were selected via $M_{\mathrm{FoF}}\geq10^{15}\,\mathrm{M}_{\astrosun}$, compared to the BAHAMAS sample.
Using the BAHAMAS sample, if we restrict our analysis to clusters with similar values of $M_{\mathrm{500,sim}}\,/\,M_{\mathrm{FoF}}$ we find that these systems are just forming and very dynamically disturbed and that the amplitude of the scatter in estimated mass increases.
Therefore, the selection of the MACSIS sample likely explains why the scatter in estimated mass is larger at fixed mass relative to the BAHAMAS sample.
We find that the scatter for the MACSIS clusters is generally flat, though there is evidence that the scatter decreases with mass before increasing again as the sample statistics become small.

Accounting for selection effects, the simulated samples overall produce a similar level of scatter in estimated mass at fixed mass, suggesting that it is insensitive to the implementation of subgrid physics.
As the numerical resolution decreases from that at which the model was calibrated, we find that the scatter in estimated mass increases marginally.
Overall, the amplitude of the hydrostatic bias we recover appears insensitive to the numerical choices, yielding a similar trend with mass and level of scatter.
Additionally, the bias recovered shows good broad agreement with the bias measured observationally using weak lensing masses as a proxy for true mass.

\section{Spherically symmetric haloes}
\label{sec:ssh}
The other assumption required for many cluster mass estimation techniques is that galaxy clusters are spherically symmetric.
Often in cluster studies \citep[e.g.][]{Mantz2015} the most regular systems are selected because it is believed their three-dimensional properties can be recovered with reduced systematic uncertainty, which reduces the systematic uncertainty on the mass estimate.
We now explore this assumption using the mass tensor.
Specifically, we measure how spherical a cluster is via its sphericity, with an idealized spherical halo yielding $s=1$.
As a measure of how reliable the mass estimate is we compute the relative standard deviation (RSD), also known as the coefficient of variation, for the different projections of each cluster
\begin{equation}
 RSD = \frac{1}{\langle M_{\mathrm{500,x{-}ray}}\rangle}\sqrt{\frac{\sum(M_{\mathrm{500,x{-}ray}}-\langle M_{\mathrm{500,x{-}ray}}\rangle)^{2}}{N-1}}\;,
\end{equation}
where the sum runs over all of the projections and the angle brackets denote the mean value of the estimated mass. 
If the profiles of more regular clusters can be recovered with less systematic uncertainty, then one would expect the RSD value to decrease for clusters whose sphericity values are larger.

In Fig. \ref{fig:shape-mass}, we plot the estimated mass RSD as a function of sphericity for the different simulated samples.
Additionally, to explore any trend with mass we colour the symbol based on the true mass of the cluster.
Overall, all of the simulated samples in this work show the same trend: a reduction in the value of the RSD as the sphericity of the cluster increases.
The median estimated mass difference is $\Delta M_{\mathrm{500,x{-}ray}}\approx0.1$.
Therefore, our result confirms that more spherical clusters yield estimated masses with reduced systematic uncertainty.
We note that the MACSIS sample has a lower median sphericity value relative to the other simulated samples.
This is driven by the fact that the sample is more massive and, therefore, has formed more recently.
These clusters have had less time to relax since their formation and the most spherical clusters are absent from the sample.
Finally, there is no discernible trend with cluster mass for any of the samples, which likely reflects a large range of formation histories possible for a cluster at fixed mass.

Additionally, we split the simulated samples plotted in Fig. \ref{fig:shape-mass} into relaxed and unrelaxed via two criteria.
In the top row, we use the energy ratio $E_{\mathrm{ratio}}$, a theoretical, $3$D aperture criterion that measures the ratio of kinetic energy to thermal energy for gas cells/particles within a spherical aperture, to define a relaxed subset.
In the bottom row, we classify the clusters using the centroid shift, an observational $2$D aperture criterion that measures how much the centroid of the X-ray surface brightness shifts when the aperture used to compute it changes.
The definitions of the energy ratio and centroid shift and how we compute them are presented in Sections \ref{sec:simprops} and \ref{sec:csc}, respectively.
Relaxed (unrelaxed) clusters are denoted by filled (open) symbols.
Observationally, relaxed clusters play a prominent role in studies of cluster astrophysics, scaling relations and cosmology because they are thought to be the most regular systems with minimal systematic uncertainties.
For both criteria used in this work we find the same result, selecting relaxed clusters removes those objects with the largest RSD values.
Interestingly, the selection of relaxed clusters does not lead to any significant change in the median sphericity value of the sample.
Finally, we note that although both relaxation criteria remove the largest RSD values they do not select the same clusters, with $48$, $49$, $54$, $72$ and $19$ per cent of those clusters classified as relaxed by the energy ratio appearing in the subset defined as relaxed by the centroid shift for TNG300 level 1, level 2, level 3, BAHAMAS and MACSIS, respectively.
A detailed study of many common relaxation criteria, both theoretical and observational, is presented in Cao et al. (in prep.).

\section{Origin of mass dependent bias}
\label{sec:omdb}
We now seek to understand why the hydrostatic bias becomes mass-dependent for profiles derived from the synthetic X-ray images.
As shown in eq. \ref{eq:mest}, the hydrostatic mass estimate depends linearly on the temperature measured at the radius of interest, i.e. $r_{\mathrm{500,x{-}ray}}$.
Therefore, we use the estimated mass to derive the radial point of interest and then extract both the mass-weighted and spectroscopic temperatures.
In practice, we interpolate the radial temperature profiles to obtain the temperature estimates at the radius of interest.

\begin{figure}
 \centering
 \includegraphics[width=\columnwidth,keepaspectratio=True]{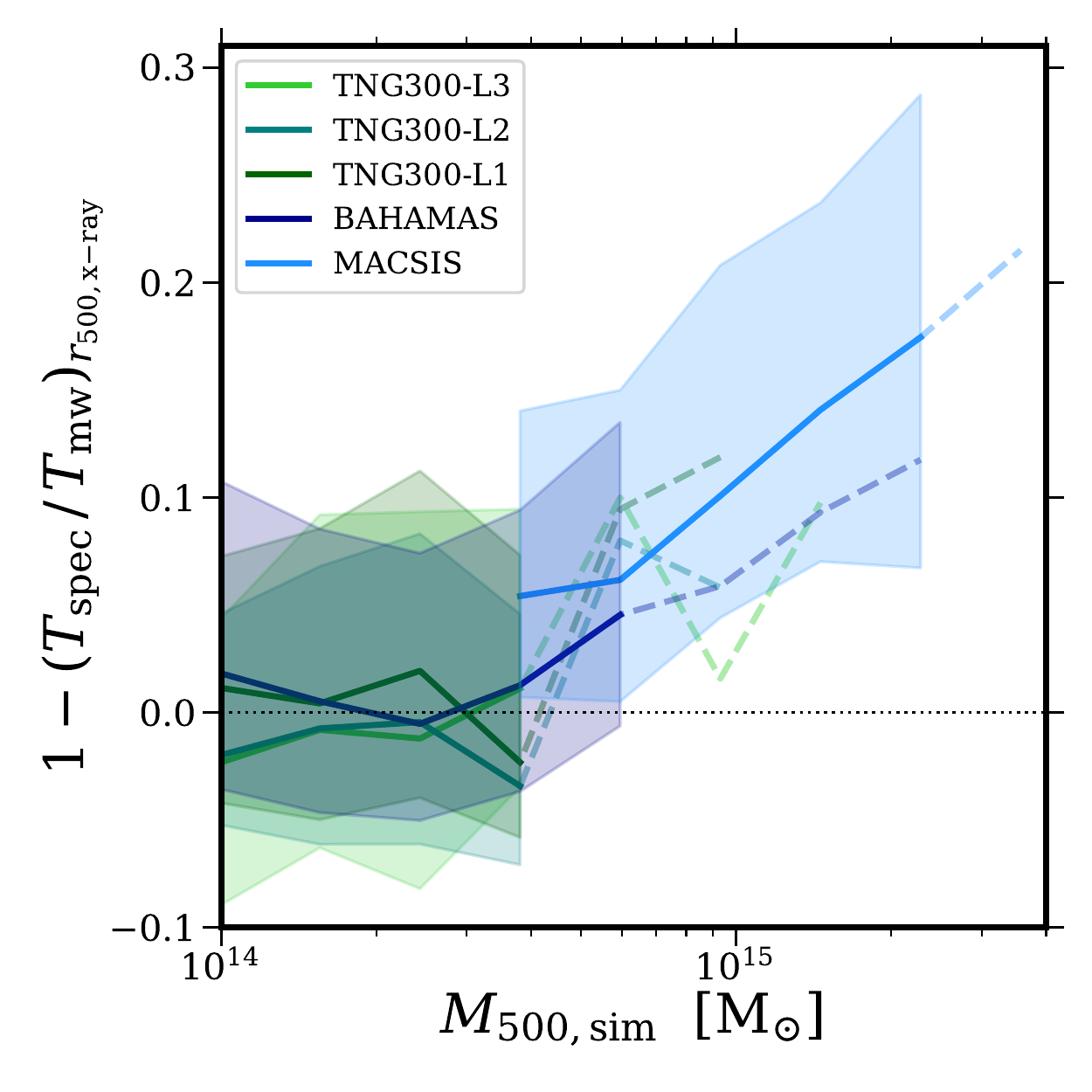}
 \caption{Median spectroscopic temperature bias relative to the mass-weighted temperature measured at $r_{\mathrm{500,x{-}ray}}$ as a function of true mass. Line styles are the same as Fig. \ref{fig:Mbias}. For low-mass clusters $(M_{\mathrm{500,sim}}<4\times10^{14}\,\mathrm{M}_{\astrosun})$ the two temperature estimates are consistent within the scatter of the samples. However, at higher masses, the spectroscopic temperature estimate consistently underpredicts the mass-weighted estimate for all simulation samples. The amplitude of this bias is sufficient to account for the mass dependence of the hydrostatic mass bias.}
 \label{fig:Tbias}
\end{figure}

In Fig. \ref{fig:Tbias}, we plot the median bias of the spectroscopic temperature relative to the mass-weighted temperature as a function of the cluster's true mass for all simulated samples.
For low-mass clusters $(M_{\mathrm{500,sim}}<4\times10^{14}\,\mathrm{M}_{\astrosun})$, the spectroscopic temperature is consistent with the mass-weighted temperature, though there is significant scatter in the samples.
We find that the numerical resolution and the subgrid physics implementation does not affect the recovered temperatures, with the TNG300 level 1, 2, 3 and BAHAMAS samples all yielding similar median values around zero.
The scatter in the bias is slightly larger for all TNG samples.
However, as the mass of the cluster increases, we find that the spectroscopic temperature is increasingly biased low relative to the mass-weighted estimate.
This result is dominated by the MACSIS cluster sample but the other simulated samples also show the spectroscopic measurement is biased low, though we are limited by small sample statistics.
For the most massive clusters $(M_{\mathrm{500,sim}}\geq2\times10^{15}\,\mathrm{M}_{\astrosun})$ the median bias reaches $0.2$, i.e. the spectroscopic temperature is only $80$ per cent of the mass-weighted temperature, and the scatter reaches values close to $0.3$.
Given the linear dependence of the hydrostatic mass on the temperature at the radius at which the estimate is made, the increasingly bias spectroscopic measurement is sufficient to explain the apparent mass dependence.
For massive clusters, the mass bias can be explained as the sum of the bias induced by assuming the cluster is in hydrostatic equilibrium $(b=0.1)$ and the bias produced by the spectroscopic temperature being lower than the mass-weighted temperature $(b=0.2)$, yielding a total bias of $b=0.3$.
Those clusters with the largest scatter would agree with the hydrostatic bias found for the most massive clusters in \citet{Pearce2020}.

\begin{figure*}
 \centering
 \includegraphics[width=\textwidth,keepaspectratio=True]{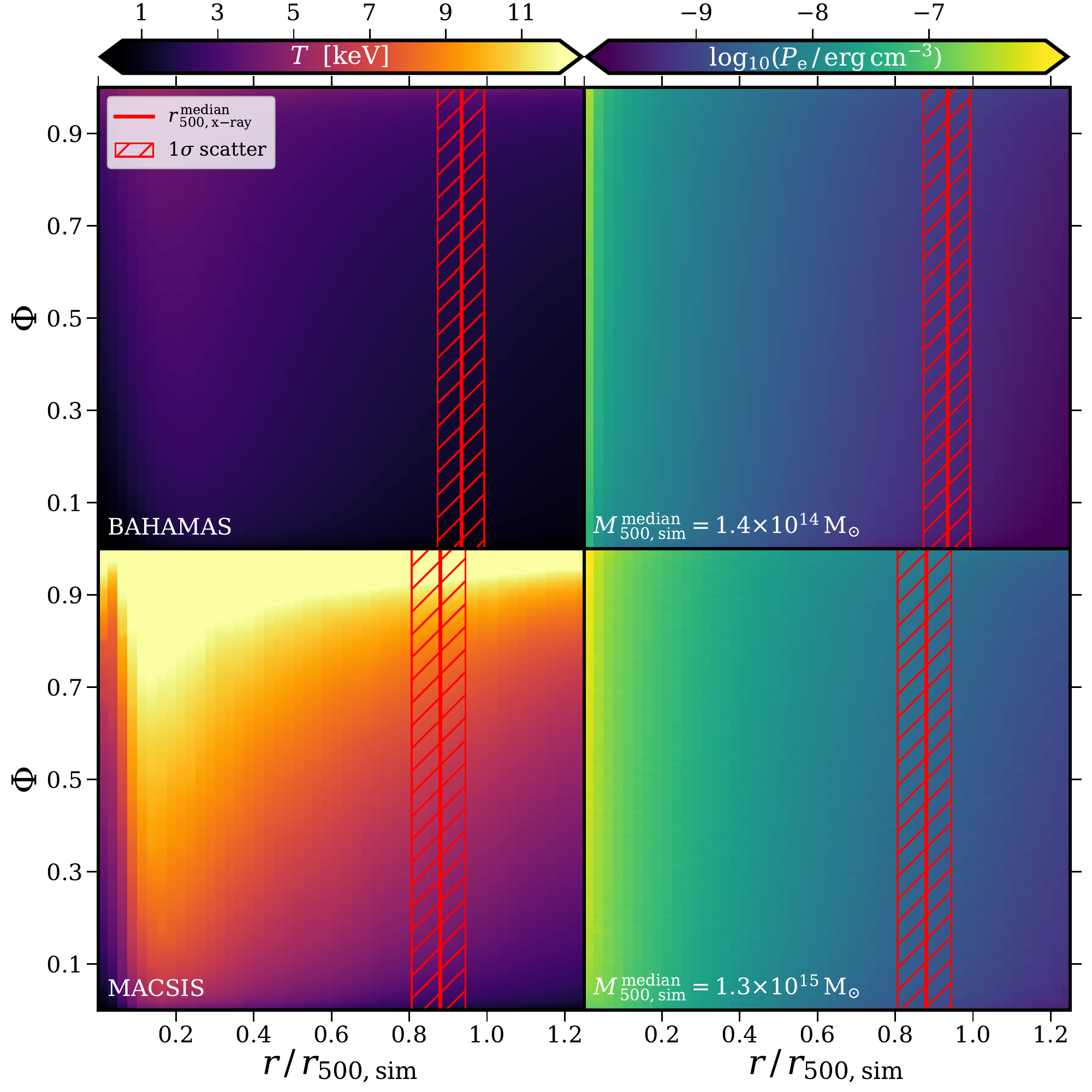}
 \caption{Temperature (left) and pressure (right) distributions for the BAHAMAS sample (top) and those clusters in the MACSIS sample with $M_{\mathrm{500,sim}}\geq10^{15}\,\mathrm{M}_{\astrosun}$ (bottom). The red line denotes the median $r_{\mathrm{500,x{-}ray}}\,/\,r_{\mathrm{500,sim}}$ value and the hashed region shows the $1\sigma$ scatter in the ratio. The MACSIS selection removes those clusters impacted by sample selection effects. The pressure distribution is ordered by the temperature of the gas. Despite the large temperature range in the MACSIS sample, the gas has a significantly smaller range in pressure. Therefore, the cooler gas must have a higher density and a larger relative contribution to the spectroscopic temperature, due to the quadratic density weighting of the emission process.}
 \label{fig:Tdist}
\end{figure*}

To understand why the spectroscopic temperature estimate is biased low relative to the mass-weighted temperature, we now explore the distribution of gas temperatures and pressures as a function of radius for two subsets of simulated clusters.
We select the entire BAHAMAS sample, noting that similar results are obtained for all resolution levels of IllustrisTNG, and all MACSIS clusters with a mass $M_{\mathrm{500,sim}}\geq\times10^{15}\,\mathrm{M}_{\astrosun}$.
The MACSIS cluster subset ensures that any impact of the sample selection function is minimized.
For each cluster, the gas cells/particles are binned into $50$ linearly spaced radial bins in the range $0-1.25\,r_{\mathrm{500,sim}}$.
Within each radial bin, the mass-weighted cumulative distribution function of the temperature is computed and stored at $200$ percentile points, i.e. increments of $0.5$.
Additionally, the volume-weighted pressure distribution is computed in each radial bin for the cells/particles still ordered by their temperature.
For both samples, the value of the median percentile in every pixel is then computed.
At this stage, it is important to account for the fact the more massive clusters are hotter and have larger pressures due to their deep potential wells.
The mass dependence is removed from the temperature and pressure distributions by dividing them by the virial temperature, $k_{\mathrm{B}}T_{500}$, and pressure, $P_{500}$, expected for a system with mass $M_{\mathrm{500,sim}}$.
These are calculated via
\begin{equation}
 k_{\mathrm{B}}T_{500} = \frac{GM_{\mathrm{500,sim}}\mu m_{\mathrm{p}}}{2r_{\mathrm{500,sim}}}\:,
\end{equation}
and
\begin{equation}
 P_{500} = 500f_{\mathrm{b}}k_{\mathrm{B}}T_{500}\frac{\rho_{\mathrm{crit}}}{\mu m_{\mathrm{p}}}\:,
\end{equation}
where $f_{\mathrm{b}}$ is the universal baryon fraction and $\rho_{\mathrm{crit}}$ is the critical density of the Universe at $z=0.1$.
For both samples, the median value in every pixel is then computed and we reintroduce the mass dependence by multiplying back through by the median values of $k_{\mathrm{B}}T_{500}$ and $P_{500}$.

Fig. \ref{fig:Tdist} shows the temperature and pressure distributions as a function of radius for the two samples.
As expected, for BAHAMAS the average temperature decreases with radius.
A similar result is seen for the IllustrisTNG samples, as seen in \citet[e.g.][]{Barnes2018,Barnes2019}.
At $r_{\mathrm{500,x{-}ray}}$, all of the gas resides within a small temperature range $(1.1-2.8\,\mathrm{keV})$. 
The MACSIS samples yields the same radial decline in temperature, but, as expected for a more massive sample, the temperature normalization is larger.
At any radius the MACSIS sample has a significantly wider temperature distribution, with the gas temperatures in the $3.1-13.1\,\mathrm{keV}$ at $r_{\mathrm{500,x{-}ray}}$.
Examining the gas pressure we find both samples yield distributions with a similar width, the pressure varies by a factor $2.5$ at $r_{\mathrm{500,x{-}ray}}$.
For the MACSIS sample the width of the pressure distribution is significantly smaller than the temperature distribution.
Given that $P=nk_{\mathrm{B}}T$, the wider temperature distribution in the MACSIS sample implies a wider distribution of densities, with cooler gas having a linearly higher relative density.
This difference in density is further amplified by the $n^{2}$ dependency of the X-ray emission mechanism.
Therefore, the cooler gas is disproportionately represented in the X-ray emission and the spectroscopic fitting yields a lower temperature than the mass-weighted temperature estimate for the most massive clusters.

This leads to the question of why this lower entropy gas has not sunk to centre of these massive clusters.
The likely answer is that these haloes are only just collapsing by the current epoch, and this infalling, colder gas has not had the opportunity to settle closer to the cluster centre.
However, examining the history and fate of this gas is beyond the scope of this study and we leave it to future work.

The mass dependence of the hydrostatic bias found in this study is driven by the result of fitting a single temperature plasma model to an X-ray spectrum that is the sum of gas at a range of temperatures.
Given the unimodal nature of temperature distribution for the massive clusters present in the MACSIS sample, it is not obvious that fitting a two (or more) component model to the spectrum is the correct approach to removing this bias.
The recent X-COP study found that the pressure profiles derived from X-ray data were in good agreement with the pressure profiles from SZ observations \citep{Ghirardini2019}.
Due to the different emission mechanism, observations of the SZ signal should yield a temperature estimate that is more closely aligned with the expected mass-weighted temperature.
Therefore, they found that the mass-weighted and spectroscopic temperature estimates were in good agreement with each other.
This study agrees with the X-COP results.
The median mass of the X-COP sample is $5\times10^{14}\,\mathrm{M}_{\astrosun}$ \citep{Eckert2017}, where we find that the median spectroscopic temperature is biased $3$ per cent lower than the mass-weighted temperature and the scatter encompasses an unbiased result.
Therefore, the simulated spectroscopic and mass-weighted pressure profiles will agree with each other.
The limitation of the X-COP study is the small number of low-mass objects $(12)$ studied.
The next generation of detailed surveys that sample a larger number of clusters, with significant statistics for very massive clusters (like the \textit{XMM} Heritage cluster project), have the potential to fully characterize the amplitude of the hydrostatic bias as a function of cluster mass.

\section{Conclusions}
\label{sec:concs}
In this work, we have introduced the multiwavelength analysis framework \textsc{Mock-X}.
The framework is agnostic to the chosen simulation, generating synthetic multiwavelength images of simulated haloes.
We have generated synthetic images for $18,756$ projections of mass-limited $(M_{\mathrm{500,sim}}>10^{14}\,\mathrm{M}_{\astrosun})$ samples from all three resolution levels of the IllustrisTNG $(300)^{3}\,\mathrm{Mpc}^{3}$ volume, the reference \textit{Planck} BAHAMAS volume and the MACSIS simulations.
We have used the synthetic Chandra-\textit{like} X-ray images to explore the bias induced by estimating cluster masses assuming that they are in hydrostatic equilibrium.
The mass estimates are derived from the thermodynamic profiles extracted directly from the synthetic images, with our chosen approach following common observational methods (Section \ref{sec:methods}).
Our main results are as follows:
\begin{itemize}
 \item Hydrostatic mass estimates derived from mass-weighted simulation profiles yield a bias of $b=0.11-0.15$ with significant scatter in the population (Fig. \ref{fig:Mbias}).
 The choice of hydrodynamical method or subgrid physics implementation has a negligible impact on the recovered bias.
 However, when masses are estimated using the thermodynamic profiles derived from synthetic X-ray images we find that the bias shows a significantly stronger mass dependence. The bias increases from $b=0.1$ at $10^{14}\,\mathrm{M}_{\astrosun}$ to $b=0.3$ at $2\times10^{15}\,\mathrm{M}_{\astrosun}$.
 \item We compare the bias recovered from synthetic X-ray images to a collection of hydrostatic mass bias estimates from \citet{Miyatake2019} and references therein.
 The observations use mass estimates derived from weak lensing as a proxy for the true cluster mass.
 We find that the amplitude of the hydrostatic bias yielded by the simulations is in excellent agreement with observed bias amplitude (Fig. \ref{fig:Obs}).
 Additionally, the sample variance of the observations is in good agreement with the scatter in mass bias found for the simulated samples.
 \item We examine the impact of projection on the recovered hydrostatic mass estimate (Fig. \ref{fig:MBproj}) and find it has a negligible impact on the recovered mass.
 The amplitude and scatter in recovered mass is consistent for all projections and numerical samples.
 \item The scatter in estimated mass is roughly constant as a function of mass (Fig. \ref{fig:Mscatter}).
 There is some evidence it decreases slightly with increasing cluster mass, but the scatter increases again as the sample statistics become small.
 Additionally, we find some evidence the mass scatter increases as the resolution of the TNG simulation decreases.
 The BAHAMAS and TNG300 level 1 samples yield consistent scatter, suggesting it is independent of chosen numerical method or subgrid physics.
 The MACSIS sample has a higher scatter at fixed mass, but this is driven by the method used to select the sample (i.e. via $M_{\mathrm{FoF}}$).
 \item We examined the relative standard deviation of the estimated mass as a function of halo sphericity (Fig. \ref{fig:shape-mass}), finding the expected result that the mass is more reliably estimated for those clusters that are more spherical.
 We then select relaxed clusters using the theoretical energy ratio criterion and the observational centroid shift criterion.
 Although the two criteria select those clusters whose mass is more consistently recovered, neither preferentially select more spherical haloes nor do they select consistent subsets of relaxed clusters.
 \item We then explored the origin of the mass dependence of the hydrostatic bias by exploring the bias induced by measuring temperatures spectroscopically, rather than mass-weighted, at the radius of interest $r_{\mathrm{500,x{-}ray}}$ (Fig. \ref{fig:Tbias}).
 At low mass $(M_{\mathrm{500,sim}}=10^{14}\,\mathrm{M}_{\astrosun})$ we find that the spectroscopically estimated temperature is consistent with the mass-weighted temperature.
 However, for high-mass clusters $(M_{\mathrm{500,sim}}=2\times10^{15}\,\mathrm{M}_{\astrosun})$ the spectroscopic temperature is only $80$ per cent of the mass-weighted temperature.
 \item We then examined the temperature distribution of the gas as a function of radius (Fig. \ref{fig:Tdist}) and find that for low-mass clusters there is a narrow range $(\sim2\,\mathrm{keV})$ of temperatures at a given radius. 
 For high-mass clusters, the width of the temperature distribution increases to $>10\,\mathrm{keV}$.
 However, at any given radius we find that the width of the pressure distribution is significantly smaller.
 Given the $P=nk_{\mathrm{B}}T$, this implies that the cooler gas in massive clusters is denser than the hot gas.
 Due to the quadratic density dependence of the X-ray emission, this implies that the cold gas has a greater weighting in the spectroscopic temperature fit that decreases the recovered temperature for more massive clusters.
 Given the hydrostatic mass estimate depends linearly on temperature (equation \ref{eq:mest}), this result explains why the hydrostatic bias is mass-dependent for thermodynamic profiles derived from synthetic X-ray images.
 We left the origin and fate of this colder gas to a future study.
\end{itemize}
Despite the mass dependent bias found in this study, the results presented in this work are consistent with current observational results.
For example, at the median mass of the X-COP sample we find that the spectroscopic temperature is statistically consistent with the mass-weighted temperature and the X-ray and SZ pressure profiles should agree.
Both future simulations and observations will be important tests of this result.
Larger volume numerical simulations with a different hydrostatic method \citep[e.g.][]{Cui2018} that varies the subgrid physics implementation \citep[e.g.][]{Kannan2017,Barnes2019} will be important to further assess the impact of numerical choices on the hydrostatic bias.
Future observational projects, such as the X-COP successor the XMM heritage cluster project, that will yield detailed X-ray observations of more massive clusters that can confirm the presence of a mass-dependent hydrostatic bias.

\section*{Acknowledgements}
We thank Ian McCarthy and Florian Ruppin for useful discussions and insightful comments. 
MV and DB acknowledges support through an MIT RSC award, a Kavli Research Investment Fund, NASA ATP grant NNX17AG29G, and NSF grants AST-1814053, AST-1814259 and AST-1909831.
Part of the analysis performed in this work was run on the Harvard Odyssey clusters and the Comet HPC resource at San Diego Supercomputer Center as part of XSEDE through TG-AST180025.
Additionally, this work used the DiRAC@Durham facility managed by the Institute for Computational Cosmology on behalf of the STFC DiRAC HPC Facility (www.dirac.ac.uk).
The equipment was funded by BEIS capital funding via STFC capital grants ST/K00042X/1, ST/P002293/1, ST/R002371/1 and ST/S002502/1, Durham University and STFC operations grant ST/R000832/1. DiRAC is part of the National e-Infrastructure.

\bibliographystyle{mnras}
\bibliography{main}

\bsp	
\label{lastpage}
\end{document}